\documentclass[%
reprint,
superscriptaddress,
 amsmath,amssymb,
pra,
floatfix,
 array,
]{revtex4-1}

\usepackage{graphicx}
\usepackage{dcolumn}
\usepackage{bm}
\usepackage{amsthm}
\usepackage[caption=false]{subfig}
\usepackage[usenames]{color}

\usepackage{braket}

\begin{document}


\title{Analysis of Quantum Network Coding for Realistic Repeater Networks}

\author{Takahiko Satoh}
 \email{satoh@sfc.wide.ad.jp}
\affiliation{
Graduate School of Media and Governance, 5322 Endo, Keio University, Fujisawa,
Kanagawa 252-0882, Japan}
\author{Kaori Ishizaki}
 \email{kaori@sfc.wide.ad.jp}
\altaffiliation[Current address: ]{HAL Laboratories}
\affiliation{
Graduate School of Media and Governance, 5322 Endo,  Keio University, Fujisawa,
Kanagawa 252-0882, Japan}
\author{Shota Nagayama}
 \email{kurosagi@sfc.wide.ad.jp}
\affiliation{
Graduate School of Media and Governance, 5322 Endo,  Keio University, Fujisawa,
Kanagawa 252-0882, Japan}
\author{Rodney Van Meter}
 \email{rdv@sfc.wide.ad.jp}
\affiliation{
Graduate School of Media and Governance, 5322 Endo,  Keio University, Fujisawa,
Kanagawa 252-0882, Japan}
\affiliation{
Faculty of Environment and Information Studies,
Keio University, 5322 Endo,  Fujisawa, Kanagawa 252-0882, Japan}
\date{\today}

\begin{abstract}
Quantum repeater networks have attracted attention for
the implementation of long-distance and large-scale sharing of quantum states. 
Recently, researchers extended classical network coding, which is a
technique for throughput enhancement, into quantum 
information. The utility of quantum network coding (QNC) has been
shown under ideal conditions, but it has not been studied previously
under conditions of noise and shortage of quantum resources.
We analyzed QNC on a butterfly network, which can create end-to-end
Bell pairs at twice the rate of the standard quantum network repeater approach.
The joint fidelity of creating two Bell pairs has a small penalty for
QNC relative to entanglement swapping. It will thus be useful when we
care more about throughput than fidelity. We found that the output
fidelity drops below 0.5 when the initial Bell pairs have fidelity $F
< 0.90$, even with perfect local gates. Local gate errors have a
larger impact on quantum network coding than on entanglement swapping.
\end{abstract}

\maketitle

\section{Introduction}
Researchers are striving to produce quantum communication technology
for long-range transmission of quantum
information and sharing of distributed quantum states~\cite{Lloyd_2004,kimble2008quantum,van-meter14}.
Quantum information requires a network specialized for quantum
communication.  Quantum information may enable new functions not achievable using classical information. 
For example, quantum key distribution
creates a shared random sequence of bits between two
parties~\cite{bb84,qcb}.
Because quantum information cannot in general be measured without
disturbing the state and cannot be cloned~\cite{no_cloning}, statistical tests can prove
the absence of as eavesdropper, guaranteeing the secrecy of the bit values.
QKD technology is already realized at a commercial level for urban scale,
complex topology networks~\cite{SECOQC,Sasaki_11}. 

Besides QKD, other distributed security functions~\cite{byzantine},
general purpose distributed quantum computing and blind quantum
computing~\cite{broadbent2010measurement} 
have been proposed as uses of long distance quantum communication.
In addition, the realization of inter-continental and inter-major
city QKD is also desired.

Thus, there is a growing need for large-scale quantum networks,
but the current quantum network protocol suffers from  a
distance limit set by the probability of correctly receiving a photon
through an exponentially lossy channel and other factors.
In order to solve this problem, quantum repeaters have been 
proposed~\cite{repeater} and many of the components have
been experimentally demonstrated~\cite{duan:RevModPhys.82.1209,hucul2014modular}. 
A quantum repeater has multiple important roles:
to create and share physical entanglement pairs (Bell
pairs) between nearest neighbors over short distances,
to perform purification of Bell
pairs,
and to create one long Bell pair by connecting two entangled pairs
using entanglement swapping~\cite{repeater,Briegel_2007,VanMeter_2009_2,VanMeter_2009,Munro_2010,VanMeter_2010,entanglement_swapping}.
Long range, complex quantum networks can be realized by
arranging a number of quantum repeaters and links.
However, the cost of quantum communication per unit of quantum
information (e.g. qubit) is very high compared with 
classical communication.

\begin{figure}[htbp]
\includegraphics[width=86mm]{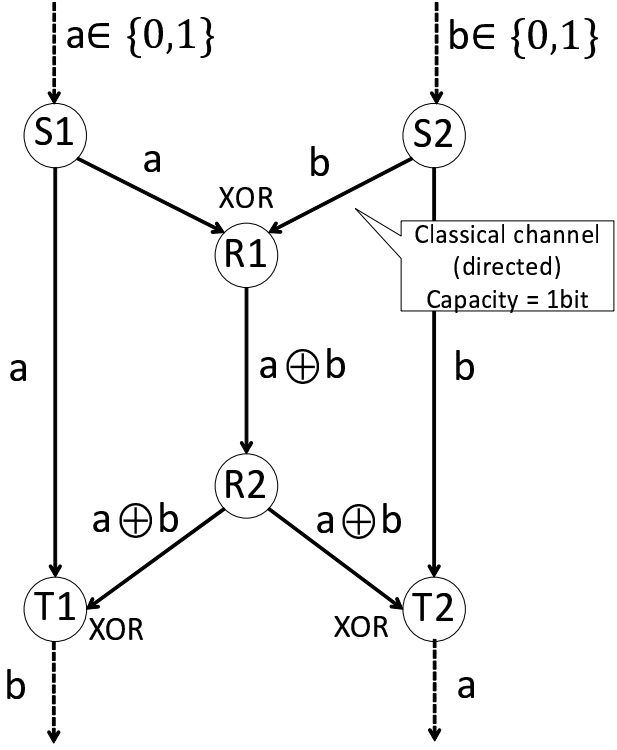}
\caption{\label{classic}
The classical network coding scheme on the butterfly network.
The problem is to send a bit of information $a$ from sender node $S1$ to target node
$T2$ and $b$ from $S2$ to $T1$ simultaneously.
It is clearly impossible to solve this problem using simply routing.
XOR operations on relay node $R1$ and target nodes $T1$, $T2$ solve
this problem.}
\end{figure}

Quantum network coding (QNC) may contribute to solving this problem.
Network coding~\cite{network_coding}  is known as a bottleneck
elimination method in classical networks. 
For example, Fig.~\ref{classic} shows simultaneous transmission
over the  directed classical butterfly network using network coding.
Two bits can be sent in one use of each link even though each
individual transmission would result in conflicts for access to
individual links.
The butterfly network is the simplest case showing a throughput bottleneck
which can be alleviated using quantum network coding.
Verifying the behavior on this graph can show that quantum network
coding can give an advantage over simple routing schemes in some
circumstances. 
It is expected that network coding also allows the same resolution in
a quantum network.
In recent years, a number of researchers have studied 
quantum network coding~\cite{quantum_coding,quantum_coding2,quantum_coding3,quantum_coding4,kobayashi_coding,kobayashi_coding2,kobayashi_coding3} .
However, all of these studies presuppose the use of pure states and perfect
local gates.
The effects of errors and resource shortages are unknown.
In this paper, we aim to determine the usefulness of quantum network
coding using mixed states.

First, we assume Pauli errors on the Bell pairs
that are our initial resources.
We investigate the error propagation in the QNC procedure and
  calculate the change of fidelities step by step in our coding scheme.
These calculations enable us
to compare  the communication efficiency between QNC
 and entanglement swapping as used in many quantum
repeater designs.
Furthermore, we calculate error thresholds for
practical QNC on the butterfly graph and find that initial resource
fidelities are required to be $F\geq 0.9$ to achieve the final
  fidelity over $0.5$.

Next, we assume Pauli errors on every CNOT
  gate, single qubit rotation, measurement, and 
quantum memory storage time step and calculate the final fidelities
using Monte Carlo simulation to assess the 
complete protocol.

The rest of this paper is organized as follows.
In Section~\ref{protocol_qnc}, we show the protocol and related
matters of quantum network coding for quantum repeaters.
In Section~\ref{error_analysis}, we present the analysis of quantum
network coding and entanglement swapping scheme in the presence of X and Z errors.
In Section~\ref{conclusion}, we conclude the discussion of this paper.

\section{Quantum network coding}
\label{protocol_qnc}
\begin{figure}[htbp]
  \includegraphics[width=86mm]{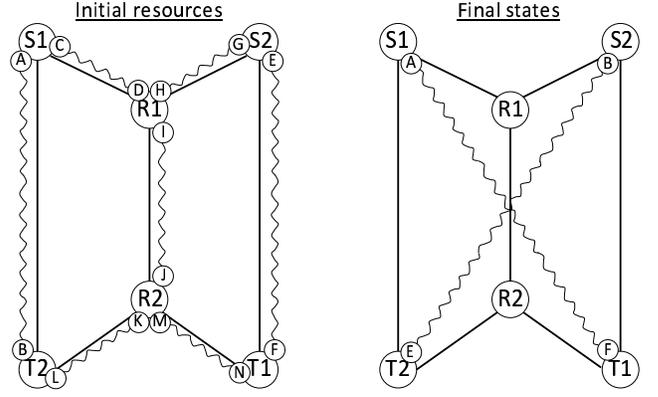}
 \caption{\label{quantumcoding_pre}Initial resources and final states for QNC. Each
   repeater node contains two or three qubits entangled with neighbors
 into Bell pairs as shown. Our goal is to establish Bell pairs between nodes
 $S1$ \& $T1$, and $S2$ \& $T2$.}
\end{figure}

Let us review the concept of quantum network coding for quantum
repeaters by examining the butterfly graph in 
Fig.~\ref{quantumcoding_pre}~\cite{repeater_coding}.
Quantum network coding, like classical network coding,
shifts the location of required communication away from the single
bottleneck link, to other links in the network, reducing demand on the
bottleneck link.  We assume that the performance of all links is the
same, and that the number of times that the most-used link must be
used to complete our operation determines our ultimate performance.
We begin with $\lvert \Psi^{+} \rangle$ Bell pairs across the seven links as shown.
In this section, we use the ket vector notation to describe the pure state
with fidelity $F=1$. In following sections, we will use the ket vector to describe 
mixed states, as discussed in the beginning  of
section~\ref{error_analysis}.

\subsection{Encoding operations}
To describe the QNC protocol, we first introduce the following three
encoding operators. They consist of CNOT gate operations,
$\hat{Z}$ basis measurement operators, and one qubit rotation based on
measurement results.
The CNOTs are  executed between a Control bit and a Bell pair, where
we designate one member of the Bell pair the Resource qubit, and the
other the Target qubit.
The Control qubit $C (C_1, C_2)$ and the Resource qubit $R (R_1, R_2)$ 
exist on the same repeater.
 An $\hat{X}$ or $\hat{Z}$ rotation is performed on the
Target qubit $T (T_1, T_2)$ if and only if the measurement result is positive.
Our operations are
\begin{eqnarray}
{\bf Con}^{C}_{R\rightarrow T}&=& \hat{X}^{S_{1}}_{T}
\hat{P}^{\pm ,S_{1}}_{\hat{Z},R} {\bf CNOT}^{(C,R)}\\ 
\label{eq_con}
{\bf Add}^{C_{1},C_{2}}_{R\rightarrow T}&=& \hat{X}^{S_{1}}_{T}
\hat{P}^{\pm ,S_{1}}_{\hat{Z},R} {\bf CNOT}^{(C_{2},R)}{\bf CNOT}^{(C_{1},R)}
\label{eq_add}
\end{eqnarray}
\begin{eqnarray}
{\bf Fanout}^{C}_{R_{1}\rightarrow T_{1},R_{2}\rightarrow T_{2}}&=&
\hat{X'}^{S_{2}}_{T_{2}} 
\hat{X}^{S_{1}}_{T_{1}} 
\hat{P'}^{\pm ,S_{2}}_{\hat{Z},R_{2}} \hat{P}^{\pm ,S_{1}}_{\hat{Z},R_{1}}\nonumber\\
&& {\bf  CNOT}^{(C,R_{2})} {\bf CNOT}^{(C,R_{1})}
\label{eq_fanout}
\end{eqnarray}
where $\hat{P}^{\pm}$ is the projective measurement operator
\begin{eqnarray}
\hat{P}^{\pm}_{\hat{X}}= \frac{1}{2}\left(\bold{1}\pm\hat{X}\right),
\hat{P}^{\pm}_{\hat{Z}}= \frac{1}{2}\left(\bold{1}\pm\hat{Z}\right),
\end{eqnarray}
$\hat{X}$ and $\hat{Z}$ are the normal Pauli operators, and $S_{1}$
and $S_{2}$ are measurement outcomes of the operator
$\hat{P}^{\pm}_{\hat{X}}$ and $\hat{P}^{\pm}_{\hat{Z}}$.

\begin{figure}[htbp]
\begin{center}
  \subfloat[${\bf  Con}^{A}_{B\rightarrow C}$]{
  \includegraphics[clip,width=0.31\columnwidth]{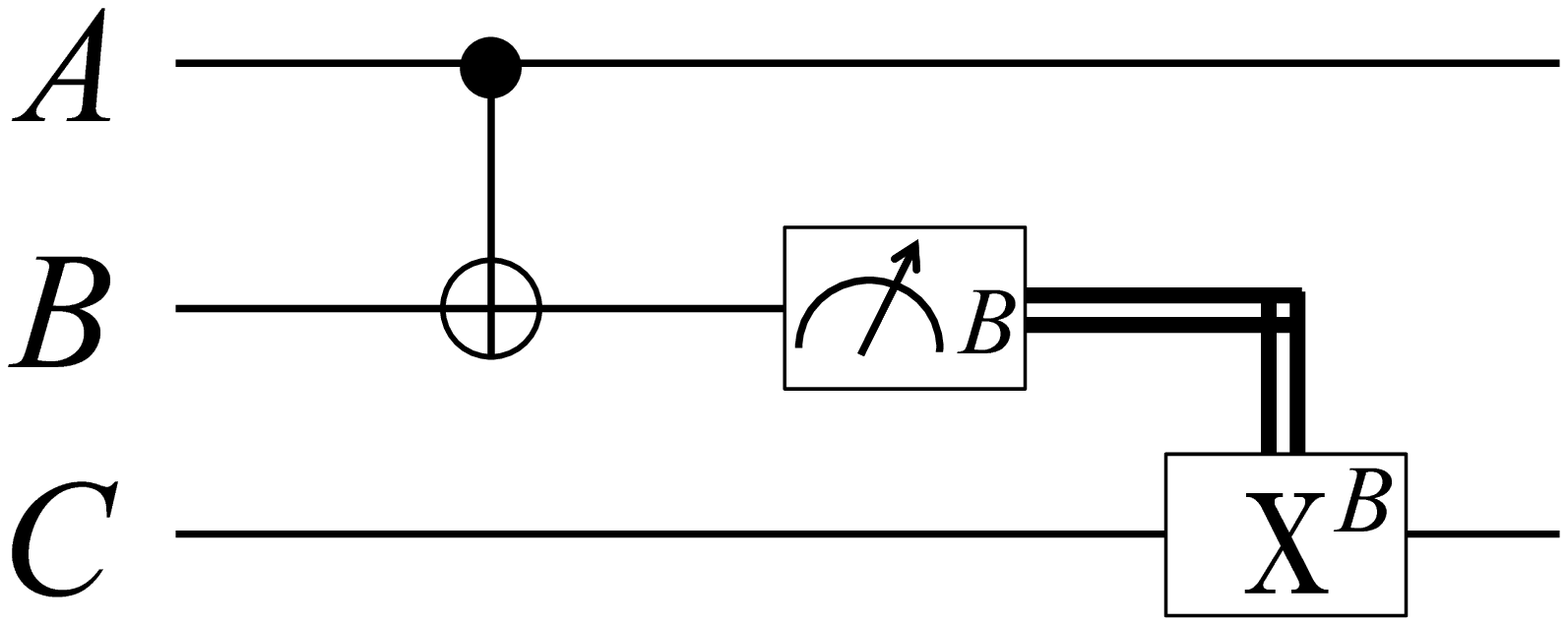}\label{connection_circuit}}
 \hfil
  \subfloat[${\bf Add}^{D,E}_{F\rightarrow G}$]{
  \includegraphics[clip,width=0.31\columnwidth]{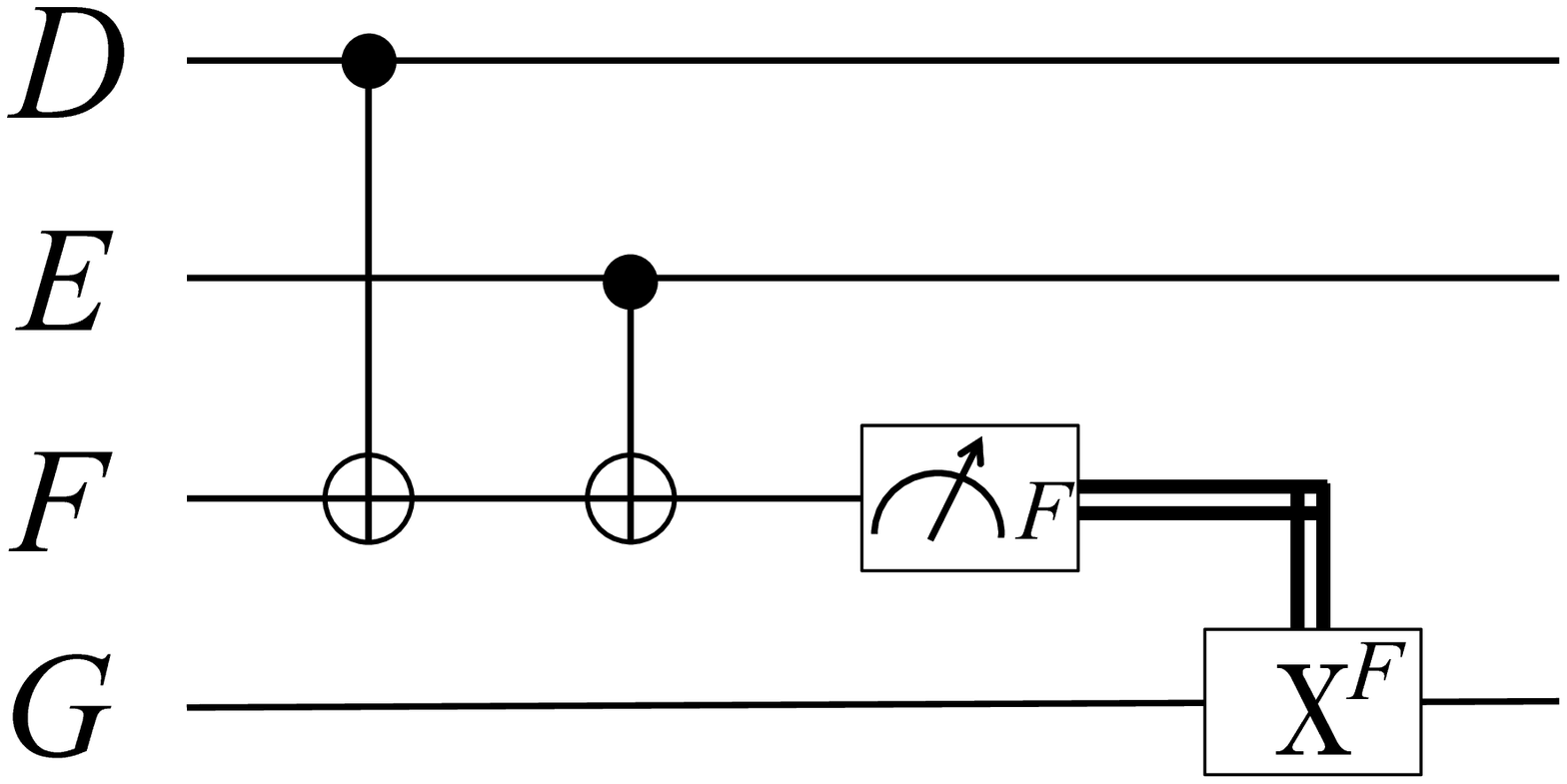}\label{add_circuit}}
 \hfil
  \subfloat[${\bf  Fanout}^{H}_{I\rightarrow J,K\rightarrow L}$]{
  \includegraphics[clip,width=0.31\columnwidth]{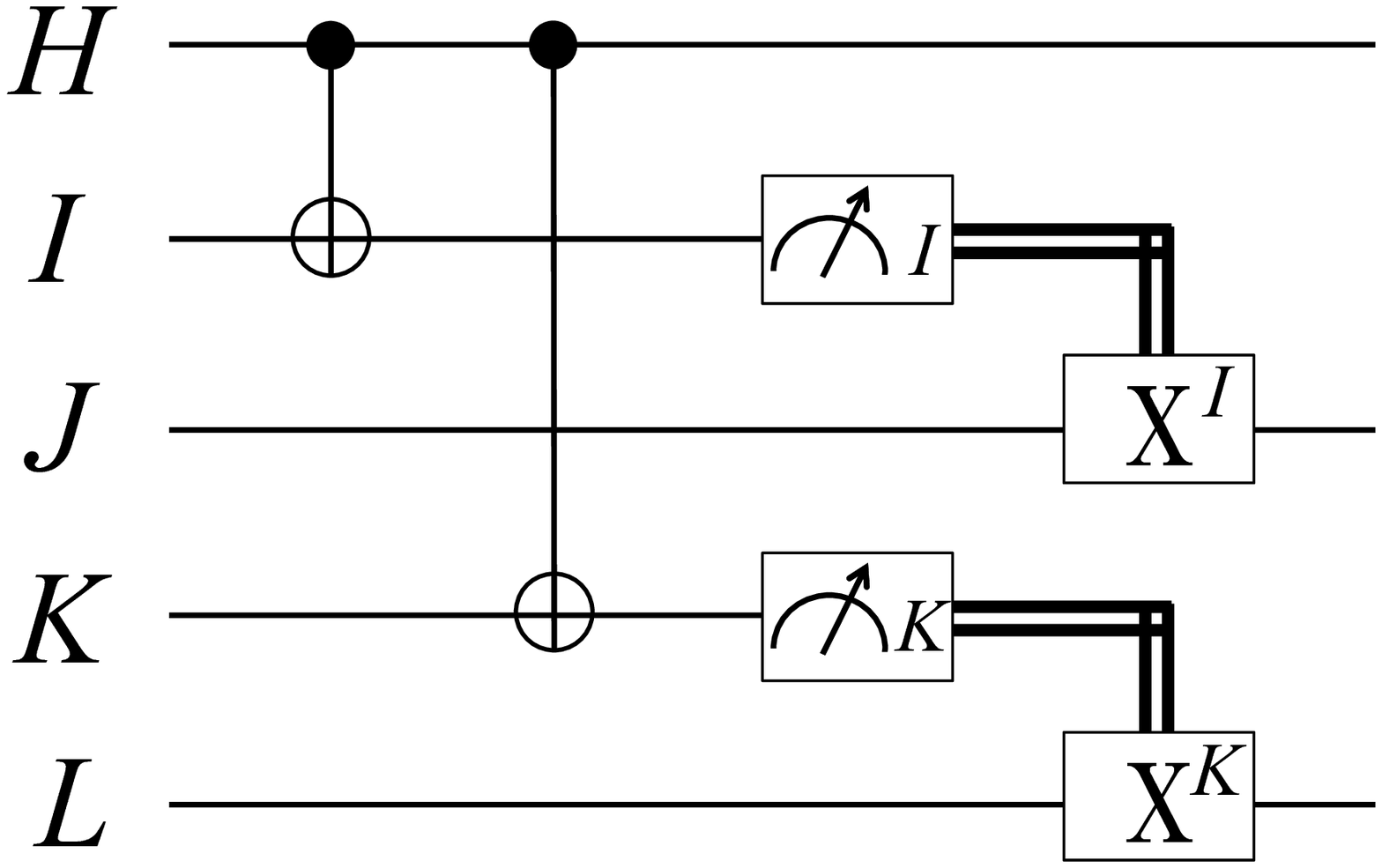}\label{fanout_circuit}}
\end{center}
 \caption{(a)\label{encoding_circuit}Connection operation between the
   qubit $A$ and the Bell pair $BC$. (b)Add
   operation between qubit $D$, $E$, and the Bell pair $FG$. (c)Fanout operation between the
   qubit $H$, and the Bell pair $IJ$ and $KL$.} 
\end{figure}

These operations correspond to the bit transfer, add and fanout operations
in a classical network coding protocol~\cite{network_coding}.
Fig.~\ref{encoding_circuit} shows quantum circuits for
  ${\bf  Con}^{A}_{B\rightarrow C}$, ${\bf Add}^{D,E}_{F\rightarrow G}$, and
  ${\bf  Fanout}^{H}_{I\rightarrow J,K\rightarrow L}$.

\subsection{Removal operations}
We also introduce the following two removal operators.
These operators are unique to quantum network coding protocols, because
we have to remove unnecessary entangled qubits before the end of
the procedure. To remove these qubits without causing changes on the remaining
system, we use $\hat{X}$ basis measurements and
feedforward operations based on the measurement results.
Our operations are
\begin{eqnarray}
{\bf Rem}_{R\rightarrow T}&=&
\hat{Z}^{S_{1}}_{T} 
\hat{P}^{\pm,S_{1}}_{\hat{X},R}\\
{\bf RemAdd}_{R\rightarrow T_{1},T_{2}}&=&
\hat{Z}^{S_{2}}_{T_{2}} 
\hat{Z}^{S_{2}}_{T_{1}} 
\hat{P}^{\pm,S_{2}}_{\hat{X},R}.
\end{eqnarray}
{\bf Rem} removes the qubits used as target qubits in the connection and
fanout operations, and  
{\bf RemAdd} removes the qubits used as target qubits in the add operations in QNC protocol.

\subsection{QNC}
Here, we introduce the protocol operator {\bf QNC} to describe the
complete procedure for
QNC. All operations in this procedure are LOCC as shown  above.
\begin{eqnarray}
{\bf QNC}\lvert \psi^{\bf{QNC}}_{init} \rangle &=&{\bf Rem}_{H\rightarrow E}{\bf
  Rem}_{D\rightarrow A}{\bf RemAdd}_{J\rightarrow {D,H}}\nonumber\\ 
&&{\bf  Rem}_{N\rightarrow J} {\bf Rem}_{L\rightarrow J} {\bf CNOT}^{(L,B)}\nonumber\\
&& {\bf CNOT}^{(N,F)}{\bf Fanout}^{J}_{K\rightarrow L, M\rightarrow N} \nonumber\\
&& {\bf Add}^{{D,H}}_{I\rightarrow J} {\bf  Con}^{E}_{G\rightarrow H} {\bf Con}^{A}_{C\rightarrow D}
\lvert \psi_{init} \rangle\\
&=& \lvert\psi^{\bf{QNC}}_{final}\rangle = \lvert \Psi^{+} \rangle_{AF} \otimes
\lvert \Psi^{+} \rangle_{BE} .
\end{eqnarray}
Here,
\begin{eqnarray}
\lvert \psi^{\bf{QNC}}_{init} \rangle &=& 
\lvert \Psi^{+} \rangle_{AB} \otimes
\lvert \Psi^{+} \rangle_{CD} \otimes
\lvert \Psi^{+} \rangle_{EF} \otimes
\lvert \Psi^{+} \rangle_{GH} \nonumber \\
&& \otimes 
\lvert \Psi^{+} \rangle_{IJ} \otimes
\lvert \Psi^{+} \rangle_{KL} \otimes
\lvert \Psi^{+} \rangle_{MN} .
\end{eqnarray}
When we perform {\bf QNC} on the seven Bell pairs, we can create two crossed Bell pairs as
a result. In this state, we can perform quantum teleportation between
repeaters in opposite corners simultaneously, as shown in Fig.~\ref{quantumcoding_pre}.
The total circuit of {\bf QNC} is shown in Fig.~\ref{circuit}.

\subsection{QNC versus entanglement swapping}
To compare this QNC protocol with the existing repeater protocols, we
also introduce the protocol operator
${\bf 2ES}$. In this procedure, we perform two entanglement swapping
operations using three Bell pairs. 
\begin{eqnarray}
{\bf 2ES}\lvert \psi_{init}^{\bf{2ES}} \rangle &=&
{\bf ES}^{(C,J)}_{(M,N)}{\bf ES}^{(C,D)}_{(I,J)}\lvert \psi_{init}^{\bf{2ES}} \rangle 
   \\
&=& \lvert \psi_{final} ^{\bf{2ES}}\rangle = \lvert \Psi^{+} \rangle_{CN}. 
\end{eqnarray}
Here,
\begin{eqnarray}
{\bf ES}^{(C,D)}_{(I,J)} &=& {\bf Rem}_{D\rightarrow C} {\bf Con}^{D}_{I\rightarrow J} \\
\lvert \psi^{\bf{2ES}}_{init} \rangle &=& 
\lvert \Psi^{+} \rangle_{CD} \otimes
\lvert \Psi^{+} \rangle_{IJ} \otimes
\lvert \Psi^{+} \rangle_{MN} .
\end{eqnarray}
Entanglement swapping between two Bell pairs can generate one long
Bell pair~\cite{entanglement_swapping}.
{\bf Rem} removes the leftover qubit for this operation.

\begin{figure}[htbp]
  \includegraphics[width=86mm]{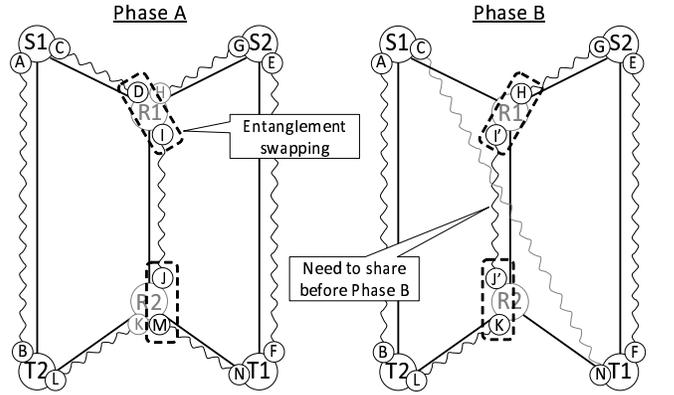}
 \caption{\label{quantumcoding_swap}Conceptual diagram of
   communication using entanglement swapping. Simultaneous execution
   of Phase A and Phase B is not possible. Re-sharing of a Bell-pair
   is needed between $R1$ and $R2$. The AB and EF Bell pairs are
   unused in this protocol.}
\end{figure}

Next, we discuss the bottleneck problem on the butterfly network.
In this case, we cannot perform ${\bf 2ES}$ two times and share two target Bell
pairs between $AF$ and $BE$
without remaking Bell pairs as shown in
Fig.~\ref{quantumcoding_swap}.
Bell pair $IJ$ is the bottleneck limiting the performance.
One approach to solving this bottleneck problem is  
link multiplexing~\cite{repeater-muxing}.
In this scheme, an approach such as time division multiplexing is proposed to solve the
bottleneck problem on a dumbbell network with few
shared Bell pairs. To compare 2ES and network coding, we adopt
this scheme.
Note that network coding generates the two goal Bell pairs while
consuming seven Bell pairs in one cycle, whereas entanglement swapping
consumes only six Bell pairs but requires two cycles because of the
resource conflict.
When we assume the time necessary to share Bell pairs between nearest
neighbor repeaters and the memory lifetime of Bell pairs are similar,
it is hard to share extra Bell pairs between bottleneck repeaters.

\section{Errors on the initial Bell pairs}
\label{error_analysis}
To elucidate the advantage of QNC, if any, we compare the
communication fidelity of QNC and 2ES.
Before tackling the more general problem including gate errors, we
investigate the propagation of X and Z errors present in the initial seven
Bell pairs in Fig.~\ref{quantumcoding_pre}. 
We define $\epsilon_{qubit,\hat{X}(\hat{Z})}$ as $\hat{X} (\hat{Z})$ rotation
error with probability $p$.
Due to the symmetry of Bell pairs, we do not need to distinguish
between an error on qubit $A$ and one on qubit $B$.
For example, we describe those two types of errors on
Bell pair $\lvert\Psi^{+}\rangle_{AB} $ as follows:
\begin{eqnarray}
\epsilon^{}_{A,\hat{X}} \lvert\Psi^{+}\rangle_{AB} &=&
F\lvert\Psi^{+}\rangle_{AB} +
(1-F)\lvert\Phi^{+}\rangle_{AB}\\
\epsilon^{}_{A,\hat{Z}} \lvert\Psi^{+}\rangle_{AB} &=&
F\lvert\Psi^{+}\rangle_{AB} + (1-F)\lvert\Psi^{-}\rangle_{AB}.
\end{eqnarray}
Here, fidelity $F = 1-p = \langle \psi \rvert \rho \lvert \psi
\rangle$ where $\lvert \psi \rangle$ is the desired pure state.
In this paper, for simplicity of representation, we retain the ket
notation even for mixed states. The above should be understood to
represent
\begin{eqnarray}
\rho &=& \sqrt{\epsilon^{}_{A,\hat{X}}} \lvert\Psi^{+}\rangle \langle
\Psi^{+} \rvert_{AB} \sqrt{\epsilon^{}_{A,\hat{X}}} \\
&=& F\lvert\Psi^{+}\rangle \langle\Psi^{+}\rvert_{AB} +
(1-F)\lvert\Phi^{+}\rangle \langle\Phi^{+}\rvert_{AB}.
\end{eqnarray}

In this section, we assume that we can perform single qubit rotation,
CNOT gate, and projective measurement perfectly with success probability
$1$. 
Gate errors will be incorporated in Sec.~\ref{incorporating errors}.

\subsection{Z errors}
Here, we discuss Z errors on our initially shared Bell pairs.
Z errors propagate via a CNOT gate from target qubit to control qubit.
\subsubsection{Connection}
\begin{figure}[htbp]
  \includegraphics[width=86mm]{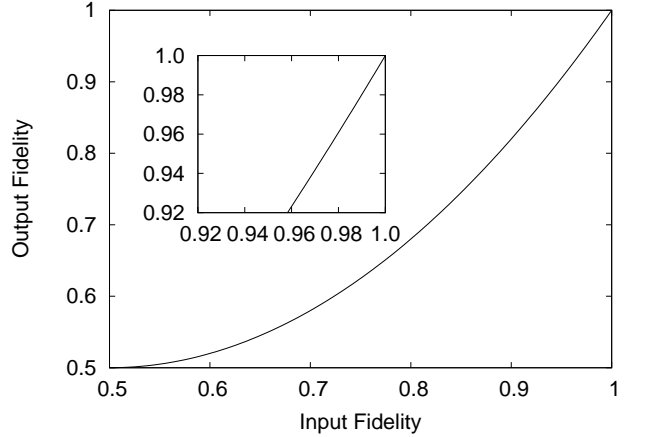}
 \caption{\label{connection-pf}Fidelity against Bell pair Z errors only during
   the Connection operation. The horizontal axis corresponds to the initial fidelity
of each Bell pair. The vertical axis corresponds to the final fidelity
of the system. Local gates are assumed to be perfect.}
\end{figure}

First, we investigate the Z error propagation in the Connection operation.
When we perform Connection ${\bf Con}^{B}_{C\rightarrow D}$ between Bell pairs
$AB$ and $CD$ with probabilistic Z errors on qubits A and C, the
Z error on measured qubit C causes a similar error on qubit B. 
Then, the initial state $\lvert \psi^{'\bf{Con}}_{init}\rangle$
 can be described as follows:
\begin{eqnarray}
\lvert \psi^{'\bf{Con}}_{init} \rangle &=&
\epsilon^{(A)}_{A,\hat{Z}} 
\lvert \Psi^{+} \rangle_{AB} \otimes \epsilon^{(C)}_{C,\hat{Z}}\lvert
\Psi^{+} \rangle_{CD}. 
\end{eqnarray}
After the Connection operation, the final state
$\epsilon^{(C)}_{B,\hat{Z}} \epsilon^{(A)}_{A,\hat{Z}}
\lvert \psi^{{\bf Con}}_{final} \rangle$ becomes 
\begin{eqnarray}
\vert000\rangle_{ABD} + \sum^{0,1}_{S_{AB}}\sum^{0,1}_{S_{CD}}  p_{S_{AB}}  p_{S_{CD}}{(-1)}^{S}\vert111\rangle_{ABD}. 
\end{eqnarray}
Here, $\epsilon^{(P)}_{Q,\hat{Z}}$ denotes a Z error on qubit $Q$
resulting from the
original Z error on qubit $P$.
$S$ is calculated as follows:
\begin{eqnarray}
S = S_{AB} + S_{CD}.
\end{eqnarray}
$S_{AB}$ and $S_{CD}$ are $1$ if the corresponding Bell pair
includes a Z error, otherwise they are $0$.
When we assume the initial fidelity of each Bell pair $F_{AB} =
F_{CD}=F$, the result is a phase flip error ($S=1$)
with probability $2F(1-F)$, otherwise $S=0$.
We show pre-operation and post-operation fidelities in
Fig.~\ref{connection-pf}. 

\subsubsection{Add}
Second, we investigate the error propagation in the Add operation.
For example, we perform ${\bf Add}^{F,H}_{I\rightarrow J}$  with three
Bell pairs $EF$, $GH$, and $IJ$.
The initial state $\lvert \psi^{'\bf{Add}}_{init}\rangle$
 can be describe as follows:
\begin{eqnarray}
\!\!\!\!\!\!\!\lvert \psi^{'\bf{Add}}_{init} \rangle \!=\! \epsilon^{(I)}_{I,\hat{Z}}
\epsilon^{(G)}_{G,\hat{Z}} \epsilon^{(E)}_{E,\hat{Z}} 
\lvert \Psi^{+} \rangle_{EF} \!\otimes\! \lvert \Psi^{+} \rangle_{GH}
\otimes \lvert \Psi^{+} \rangle_{IJ}.
\end{eqnarray}
After the Add operation, the final state
$\epsilon^{(I)}_{H,\hat{Z}} \epsilon^{(I)}_{F,\hat{Z}}
\epsilon^{(G)}_{G,\hat{Z}} \epsilon^{(E)}_{E,\hat{Z}} 
\lvert \psi^{{\bf Add}}_{final} \rangle$
becomes 
\begin{widetext}
\begin{eqnarray}
\sum^{0,1}_{S_{AB}}\!\sum^{0,1}_{S_{EF}}\!\sum^{0,1}_{S_{GH}}\!
p_{S_{AB}} p_{S_{CD}} p_{S_{EF}}
(\vert0000\rangle \!+\!
 {\!(- \! 1)}^{S_{0}}\!\vert1111\rangle)_{EFGH}\vert0\rangle_{J} 
\!+\! {\!(- \!1)}^{S_{1}}\!(\vert0011\rangle \!+\!
 {\!(- \!1)}^{S_{0}}\!\vert1100\rangle)_{EFGH}\vert1\rangle_{J}.
\end{eqnarray}
\end{widetext}
Here, $\lvert \psi^{{\bf Add}}_{final} \rangle$ corresponds to the
state in Eq.~\ref{eq_add}.
Each $S_{i}$ is calculated as follows:
\begin{eqnarray}
S_{0} &=& S_{AB}+S_{EF},\\
S_{1} &=& S_{EF}+S_{IJ}.
\end{eqnarray}
When all $S_{i}\neq 1$ where $i \in \{0,1\}$, which occurs with probability $F^3+(1-F)^3$,
then the final state is error free.

\subsubsection{Fanout}
Third, we investigate the error propagation in the Fanout operation.
When we perform ${\bf Fanout}^{J}_{K\rightarrow L, M\rightarrow N}$ with three Bell pairs
$IJ$, $KL$ and $MN$, the initial state $\lvert \psi'^{\bf{Fanout}}_{init} \rangle$
 can be describe as follows:
\begin{eqnarray}
\!\!\!\!\!\!\!\!\!\!\lvert \psi'^{\bf{Fanout}}_{init} \rangle \!\!=\!\! \epsilon^{(I)}_{I,\hat{Z}}
\epsilon^{(K)}_{K,\hat{Z}} \epsilon^{(M)}_{M,\hat{Z}} 
\!\lvert \Psi^{+} \rangle_{IJ} \!\otimes\! \lvert \Psi^{+} \rangle_{KL}
\!\otimes\! \lvert \Psi^{+} \rangle_{MN}\!.
\end{eqnarray}
After the Fanout operation, the final state
$\epsilon^{(I)}_{I,\hat{Z}} \epsilon^{(K)}_{J,\hat{Z}} \epsilon^{(M)}_{J,\hat{Z}}
\lvert \psi^{{\bf Fanout}}_{final} \rangle$
becomes
\begin{widetext}
\begin{eqnarray}
 \sum^{0,1}_{S_{IJ}}\sum^{0,1}_{S_{KL}}\sum^{0,1}_{S_{MN}}
p_{S_{IJ}}p_{S_{KL}}p_{S_{MN}}(\vert00\rangle_{LN} + {(-1)}^{S_{0}}\vert11\rangle_{LN} )
\lvert0\rangle_{J}
+{(-1)}^{S_{1}}(\vert01\rangle_{LN} + {(-1)}^{S_{0}}\vert10\rangle_{LN})\vert1\rangle_{J}.
\end{eqnarray}
\end{widetext}
Here, $\lvert \psi^{{\bf Fanout}}_{final} \rangle$ corresponds to the
state in Eq.~\ref{eq_fanout}.
Each $S_{i}$ is calculated as follows
\begin{eqnarray}
S_{0} &=& S_{EF}+S_{GH},\\
S_{1} &=& S_{GH}+S_{IJ}.
\end{eqnarray}
When all $S_{i}=0$ where  $i \in \{0,1\}$, which occurs  with
probability $F^3$, then the final state is error free.

\subsubsection{Removal and Removal-Add}
In Removal and Removal-Add operation, we perform X basis
measurement on the target qubit. When a Z error exists on the target
qubit, the measurement result flips.
Removal and Removal-Add move a Z error from the measured qubit to
the feedfoward qubit(s).
We show this error propagation below:
\begin{eqnarray}
{\bf Rem}_{R\rightarrow T}\epsilon^{(R)}_{R,\hat{Z}}\lvert \psi_{init}^{\bf{Rem}}\rangle &=& 
\epsilon^{(R)}_{T,\hat{Z}}\lvert \psi^{\bf{Rem}}_{final}\rangle ,\\
\!\!\!\!\!\!\!\!\!\!{\bf R\!e\!m\!A\!d\!d}_{R\rightarrow T_{1},T_{2}}\epsilon^{\hat{Z}}_{R}\lvert \psi^{\bf{R\!e\!m\!A\!d\!d}}_{init}\rangle &=& 
\epsilon^{(R)}_{T_{1},\hat{Z}}\epsilon^{(R)}_{T_{2},\hat{Z}}\lvert
\psi^{\bf{R\!e\!m\!A\!d\!d}}_{final}\rangle . 
\end{eqnarray}

\begin{figure}[htbp]
  \includegraphics[width=86mm]{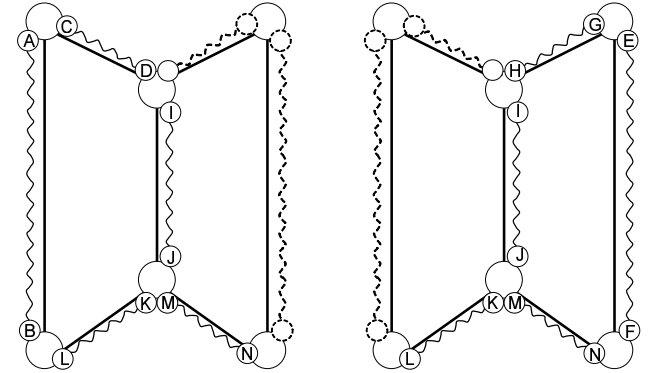}
 \caption{\label{zerror-comb}Z errors propagation. The left figure
   shows the five Bell pairs that affect the final Bell pair AF. The
   right figure shows the five Bell pairs that affect the final Bell pair BE.}
\end{figure}

To conclude the above discussion, we show the location of errors which cause Z errors on final
Bell pairs in
Fig.~\ref{zerror-comb}.

\subsubsection{Comparison}
\begin{figure}[htbp]
  \includegraphics[width=86mm]{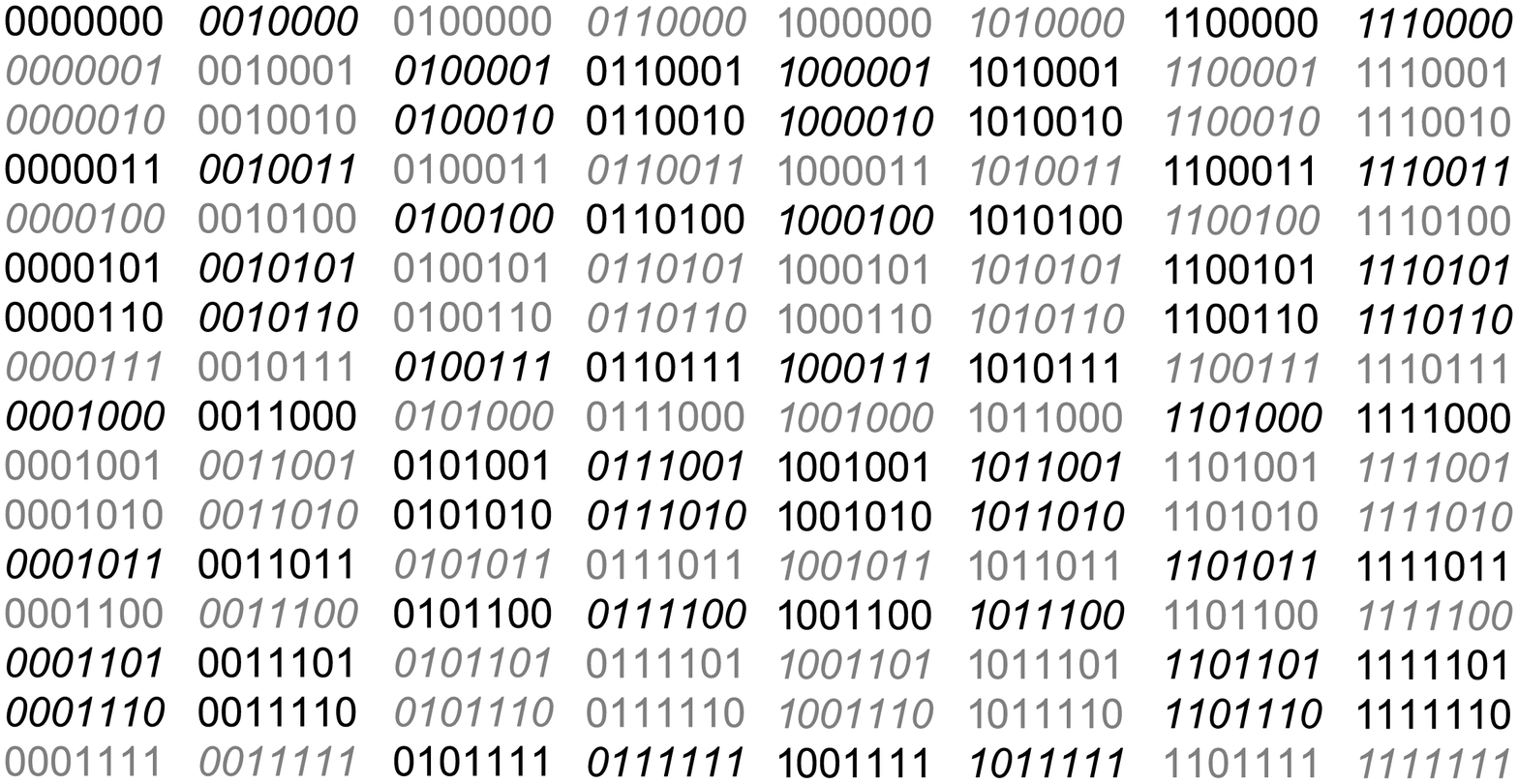}
 \caption{\label{chartZ}Chart of Z errors. Each seven-bit string
   indicates the presence $(1)$ or absence $(0)$ of a Z error on the 
Bell pairs $AB$, $CD$, .. , and $LM$, respectively. The style of the string
corresponds to the error existence on final state. Black and roman means no
error, gray(italic) means Z error on $AF$($BE$), and gray and italic means errors
on both Bell pairs.}
\end{figure}

To compare QNC and 2ES, we first calculate the final fidelity after
the complete QNC sequence.
When we assume each initial Bell pair has a Z error on one
qubit  with probability $1-F$,  the initial state $\lvert
\psi_{init}^{'\bf{QNC}}\rangle$ and final state  $\lvert
\psi_{final}^{'\bf{QNC}}\rangle$ become
\begin{eqnarray}
\!\!\!\!\!\lvert \psi_{init}^{'\bf{QNC}}\rangle &=&
\epsilon^{(M)}_{M,\hat{Z}} \epsilon^{(K)}_{K,\hat{Z}}
\epsilon^{(I)}_{I,\hat{Z}} \epsilon^{(G)}_{G,\hat{Z}}
\epsilon^{(E)}_{E,\hat{Z}} \epsilon^{(C)}_{C,\hat{Z}}
\epsilon^{(A)}_{A,\hat{Z}}\lvert \psi_{init}^{\bf{QNC}}\rangle ,\\
\!\!\!\!\!\lvert \psi_{final}^{'\bf{QNC}}\rangle &=&
\sum_{m}^{0,1}\sum_{n}^{0,1} P_{m,n} \hat{Z}^{m}_{A}
  \hat{Z}^{n}_{B} \lvert \psi_{final}^{\bf{QNC}}\rangle .
\end{eqnarray}
where $m$ and $n$ are the absence $(0)$ or presence $(1)$ of Z errors
on the final $AF$ and $BE$ Bell pairs, respectively (or equivalently
on the $A$ and $B$ qubits after use of the Bell pairs for e.g. teleportation).
The probability of each case $P_{m,n}$ is
\begin{eqnarray}
P_{0,0} &=& F^7 + 5F^5(1-F)^2 + 12F^4(1-F)^3 \nonumber \\ 
& & + 7F^3(1-F)^4 + 4F^2(1-F)^5 + 3F(1-F)^6, \\
P_{0,1} &=&  P_{1,0}= 2F^6(1-F) + 6F^5(1-F)^2 + 8F^4(1-F)^3 \nonumber \\ 
& & + 8F^3(1-F)^4 + 6F^2(1-F)^5 + 2F(1-F)^6, \\
P_{1,1} &=& 3F^6(1-F) + 4F^5(1-F)^2 + 7F^4(1-F)^3 \nonumber\\ 
& & +12F^3(1-F)^4 + 5F^2(1-F)^5  + (1-F)^7.\label{eq_probmn}
\label{result_prob}
\end{eqnarray}
Each of the 128 combinations in Fig.~\ref{chartZ} occurs with
probability $F^{(7-w)}(1-F)^{w}$ where $w$ is the Hamming weight of
the bitstring.

Next, we calculate the final fidelity in the 2ES scheme.
When we assume each Bell pair for initial resource has a Z error on one
qubit  with probability $1-F$,  the initial state $\lvert
\psi_{init}^{'\bf{2ES}}\rangle$ and final state  $\lvert
\psi_{final}^{'\bf{2ES}}\rangle$ become as follows:
\begin{eqnarray}
\lvert \psi_{init}^{'\bf{2ES}}\rangle &=&
\epsilon^{(M)}_{M,\hat{Z}} \epsilon^{(I)}_{I,\hat{Z}} 
\epsilon^{(C)}_{C,\hat{Z}} 
\lvert \psi_{init}^{\bf{2ES}}\rangle ,\\ 
\lvert \psi_{final}^{'\bf{2ES}}\rangle &=&
\sum_{m}^{0,1} P_{m} \hat{Z}^{m}_{A}
\lvert \Psi^{+}\rangle_{CN} .
\end{eqnarray}
Here, we show the probability of each case $P_{m}$ below:
\begin{eqnarray}
P_{0} &=& 1F^3 + 3F(1-F)^2 ,\\
P_{1} &=& 3F^2(1-F) + (1-F)^3.
\end{eqnarray}

\begin{figure}[htbp]
  \includegraphics[width=86mm]{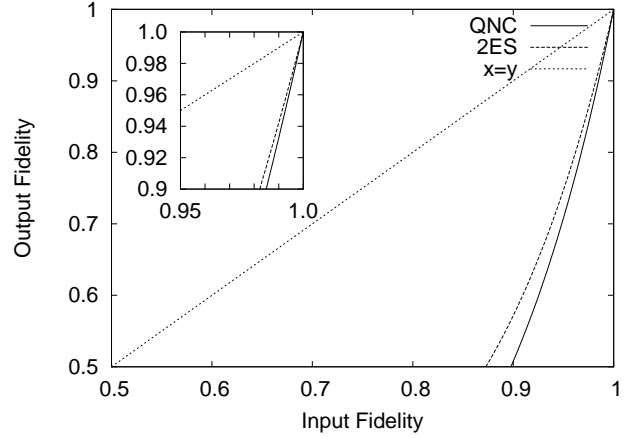}
 \caption{\label{SwappingvsQNC_Z}Comparison of Swapping and QNC with Z
   errors only. Both
   show a substantial penalty compared to the fidelity of a single
   Bell pair (the $x=y$ line).}
\end{figure}

We show the relationship between the
input fidelity and the output fidelity of our network coding protocol and
2-entanglement swapping in Fig.~\ref{SwappingvsQNC_Z}.
Here, the final state with $F_{output}<0.5$ has no practical use for
quantum communication.
When  $F_{input}\leq 0.87$, the 2ES protocol falls below $F_{out}=0.5$.
When  $F_{input}\leq 0.9$, the QNC protocol also falls below $F_{out}=0.5$.

\subsection{Classical correlation}
Next, we discuss the classical correlation between two final
Bell states. 
When we assume the input fidelity $F=0.90$, the probability of the
possible resulting states of both the AF and BE Bell pairs is
shown in Table~\ref{aleatory_Z} by the formula~(\ref{result_prob}). 
\begin{table}[htb]
\begin{center}
\begin{tabular}{c|c|c|c}
 &$\lvert \Psi_{BE}^{+} \rangle $ & $\lvert \Psi_{BE}^{-} \rangle$ &  \\
\hline
$\lvert \Psi_{AF}^{+} \rangle $ & a & b & e  \\
& 0.516 & 0.148 & 0.664 \\
\hline
$\lvert \Psi_{AF}^{-} \rangle$ & c & d & f \\
& 0.148 & 0.189 & 0.336 \\
\hline
& g & h & \\
& 0.664 & 0.336 &
\end{tabular}
\end{center}
\caption{The correlation between $\lvert \Psi_{AF} \rangle$ and
  $\lvert \Psi_{BE} \rangle$ for input fidelity $F=0.9$, Z errors only,
  and perfect local gates.}
\label{aleatory_Z}
\end{table}
The correlation coefficient is
\begin{equation}
\phi = \frac{ad-bc}{\sqrt{efgh}} \fallingdotseq 0.339.
\label{coefficient}
\end{equation}
The two output Bell pairs are unentagled using this error model but
their error probabilities are classically correlated. This correlation
is weak, despite the overlap of three Bell pairs in the left and right
halves of Fig.~\ref{zerror-comb}.

\subsection{X errors}
Next, we discuss X errors on the initially shared Bell pairs.
X errors propagate via CNOT gate from control qubit to target qubit.
\subsubsection{Connection}
First, we investigate the error propagation in Connection,
when we perform Connection ${\bf Con}^{B}_{C\rightarrow D}$ between Bell pairs
$AB$ and $CD$ with probabilistic X errors on qubits B and D. 
The initial state $\lvert \psi^{''\bf{Con}}_{init}\rangle$
 can be described as follows:
\begin{eqnarray}
\lvert \psi^{''\bf{Con}}_{init} \rangle &=&
\epsilon^{(D)}_{D,\hat{X}} \epsilon^{(B)}_{B,\hat{X}} 
\lvert \Psi^{+} \rangle_{AB} \otimes \lvert \Psi^{+} \rangle_{CD}.
\end{eqnarray}
After the Connection operation, the final state $\epsilon^{(D)}_{D,\hat{X}}  \epsilon^{(B)}_{D,\hat{X}}  \epsilon^{(B)}_{B,\hat{X}}
\lvert \psi^{{\bf Con}}_{final} \rangle$ becomes
\begin{eqnarray}
 \!\!\!\sum^{0,1}_{S_{AB}}\!\sum^{0,1}_{S_{CD}}p_{S_{AB}}p_{S_{CD}}{\hat{X}_{A}}^{S_{AB}}{\hat{X}_{D}}^{S_{CD}}{(\vert000\rangle
 +\vert111\rangle)}_{ABD}. 
\end{eqnarray}
Here, $\epsilon^{(P)}_{Q,\hat{X}}$ denotes an X error on qubit $Q$ from the
original X error on qubit $P$.

\begin{figure}[htbp]
  \includegraphics[width=86mm]{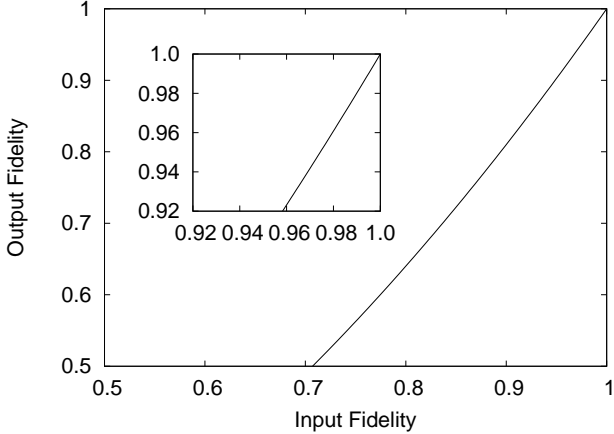}
 \caption{\label{connection-pfxa}Fidelity against X errors during Connection.}
\end{figure}

When we assume the initial Fidelity of each Bell pair $F_{AB} = F_{CD}=F$,
each $S_{i}=1$ with probability $2F(1-F)$, otherwise it is $0$.
The fidelities of the input and output
states in the Connection operation are plotted in Fig.~\ref{connection-pfxa}. 

\subsubsection{Add}
Second, we investigate the X error propagation in the Add operation.
When we perform ${\bf Add}^{F,H}_{I\rightarrow J}$ to three X error included Bell pairs $\lvert
\Psi^{+}\rangle_{EF}$, $\lvert \Psi^{+} \rangle_{GH}$, and
$\lvert\Psi^{+}\rangle_{IJ}$, the initial state $\lvert \psi''^{{\bf
    Add}}_{init} \rangle$ and  the final state $\lvert \psi''^{{\bf
    Add}}_{final} \rangle$ can be described as follows:
\begin{eqnarray}
\!\!\!\!\!\lvert \psi^{''\bf{Add}}_{init} \rangle \!=\! \epsilon^{(I)}_{I,\hat{X}}
\epsilon^{(G)}_{G,\hat{X}} \epsilon^{(E)}_{E,\hat{X}} 
\lvert \Psi^{+} \rangle_{EF} \!\!\otimes\!\! \lvert \Psi^{+} \rangle_{GH}
\!\!\otimes\!\! \lvert \Psi^{+} \rangle_{IJ}.
\end{eqnarray}
After the Add operation, the final system
$\epsilon^{(I)}_{J,\hat{X}}
\epsilon^{(G)}_{G,\hat{X}} \epsilon^{(E)}_{E,\hat{X}} 
\lvert \psi^{{\bf Add}}_{final} \rangle$
 becomes
\begin{widetext}
\begin{equation}
 \sum^{0,1}_{S_{EF}}\sum^{0,1}_{S_{GH}}\sum^{0,1}_{S_{IJ}}
p_{S_{EF}}p_{S_{GH}}p_{S_{IJ}}
\hat{X}^{S_{IJ}}_{J}\hat{X}^{S_{GH}}_{G} \hat{X}^{S_{EF}}_{E} 
((\vert0000\rangle +\vert1111\rangle)_{EFGH}\vert0\rangle_{J}
 +(\vert0011\rangle + \vert1100\rangle)_{EFGH}\vert1\rangle_{J}).
\end{equation}
\end{widetext}
When all Bell pairs' fidelities are equal, the final state's fidelity
becomes $F^3$.
\subsubsection{Fanout}
Third, we investigate the X error propagation in Fanout operation.
When we perform ${\bf Fanout}^{L}_{M\rightarrow N, O\rightarrow P}$ with three Bell pairs
$KL$, $MN$ and $OP$. Initial state $\lvert \psi''^{\bf{Fanout}}_{init} \rangle$
 can be described as follows:
\begin{eqnarray}
\!\!\!\!\lvert \psi''^{\bf{F\!a\!n\!o\!u\!t}}_{init} \rangle \!\!=\!\! \epsilon^{(K)}_{K,\hat{X}}\!
\epsilon^{(M)}_{M,\hat{X}}\! \epsilon^{(O)}_{O,\hat{X}}\! 
\lvert \!\Psi^{+}\!\rangle_{KL}\!\!\otimes\!\! \lvert\! \Psi^{+}\! \rangle_{MN}
\!\!\otimes\!\! \lvert\! \Psi^{+}\! \rangle_{OP}.
\end{eqnarray}
After Fanout, the final system
$\epsilon^{(K)}_{K,\hat{X}} \epsilon^{(M)}_{N,\hat{X}} \epsilon^{(O)}_{P,\hat{X}}
\lvert \psi^{{\bf Fanout}}_{final} \rangle$
becomes 
\begin{widetext}
\begin{eqnarray}
\sum^{0,1}_{S_{KL}}\sum^{0,1}_{S_{MN}}\sum^{0,1}_{S_{OP}}
p_{S_{KL}}p_{S_{MN}}p_{S_{OP}}\hat{X}^{S_{KL}}_{K}\hat{X}^{S_{MN}}_{N} \hat{X}^{S_{OP}}_{P}
(\vert0000\rangle + \vert1111\rangle)_{KLNP}.
\end{eqnarray}
\end{widetext}
Here, $\lvert \psi^{{\bf Fanout}}_{final} \rangle$ corresponds to the
state in Eq.~(\ref{eq_fanout}).
Each $S_{i}=1$ with probability $p$, otherwise it is $0$.
When all initial Bell pairs' fidelity are equally $F$, final state's fidelity
becomes $F^3-(1-F)^3$.

\subsubsection{Removal, Removal-Add}
In Removal and Removal-Add operations, X errors on measured qubits do
not change the measurement results. 
We describe these facts as follows:
\begin{eqnarray}
{\bf Rem}_{Q\rightarrow R}\epsilon^{(Q)}_{Q,\hat{X}}\lvert \psi_{init}\rangle &=& 
\lvert \psi_{final}\rangle ,\\
{\bf RemAdd}_{S\rightarrow T,U}\epsilon^{(S)}_{S,\hat{X}}\lvert \psi_{init}\rangle &=& 
\lvert \psi_{final}\rangle .
\end{eqnarray}

\begin{figure}[htbp]
  \includegraphics[width=86mm]{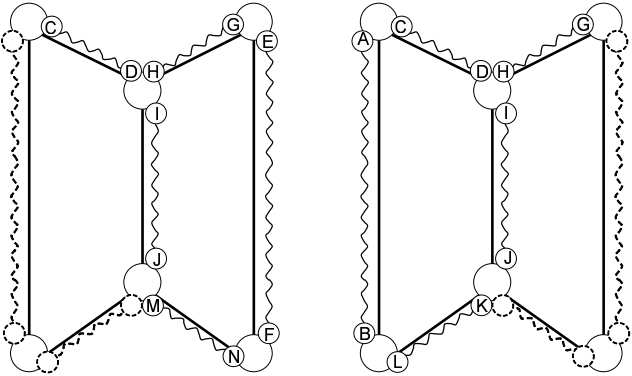}
 \caption{\label{combination}X errors propagation. The left figure
   shows the five Bell pairs that affect on the final Bell pair AF. The
   right figure shows the five Bell pairs that affect on the final Bell pair BE.}
\end{figure}

To conclude the above discussion we show the X error propagation in
Fig.~\ref{combination}.

\subsubsection{Comparison}
X error relations between the input states and the final state in the 2ES and QNC
protocols can be described as follows:
When we assume each Bell pair for initial resource has an X error on one
qubit  with probability $P_{m,n}$ are as in Eq.~\ref{eq_probmn},  the initial state $\lvert
\psi_{init}^{''\bf{QNC}}\rangle$ and final state  $\lvert
\psi_{final}^{''\bf{QNC}}\rangle$ become as follows:
\begin{eqnarray}
\!\!\!\!\lvert \psi_{init}^{''\bf{QNC}}\rangle &\!=\!&
\epsilon^{(M)}_{M,\hat{X}} \!\epsilon^{(K)}_{K,\hat{X}}
\!\epsilon^{(I)}_{I,\hat{X}}\! \epsilon^{(G)}_{G,\hat{X}}
\!\epsilon^{(E)}_{E,\hat{X}} \!\epsilon^{(C)}_{C,\hat{X}}
\!\epsilon^{(A)}_{A,\hat{X}}\!\lvert \psi_{init}^{\bf{QNC}}\rangle \\
\!\!\!\!\lvert \psi_{final}^{''\bf{QNC}}\rangle &=&
\sum_{m}^{0,1}\sum_{n}^{0,1} P_{m,n} \hat{X}^{m}_{A}
  \hat{X}^{n}_{B} \lvert \psi_{final}^{\bf{QNC}}\rangle \\
\!\!\!\!&=&\lvert \psi_{final}^{''\bf{QNC}}\rangle
\end{eqnarray}
Thus, the final fidelities of the 2ES protocol with X or Z errors are the same.
When we assume each Bell pair in our initial resource set has an X error on one
qubit  with probability $p$,
the initial state $\lvert
\psi_{init}^{''\bf{2ES}}\rangle$ and final state  $\lvert
\psi_{final}^{''\bf{2ES}}\rangle$ become as follows:
\begin{eqnarray}
\lvert \psi_{init}^{''\bf{2ES}}\rangle &=&
\epsilon^{(M)}_{M,\hat{X}} \epsilon^{(I)}_{I,\hat{X}} 
\epsilon^{(C)}_{C,\hat{X}} 
\lvert \psi_{init}^{\bf{2ES}}\rangle ,\\ 
\lvert \psi_{final}^{''\bf{2ES}}\rangle &=&
\sum_{m}^{0,1} P_{m} \hat{X}^{m}_{A}
\lvert \Psi^{+}\rangle_{CN} \nonumber\\
&=& \lvert \psi_{final}^{'\bf{2ES}}\rangle .
\end{eqnarray}

Although the fidelity is the same, the location of errors which cause
X or Z errors on the final Bell pairs are different.
As a result, the relationship between
input fidelity and output fidelity of our network coding protocol and
2-entanglement swapping are equal that of Z errors as shown in Fig.~\ref{SwappingvsQNC_Z}.

\subsection{General Pauli error model}
Finally,  we model more general errors on our initial resource Bell pairs.
 as Pauli errors occuring during CNOT gates ${\bf
  CNOT}^{(control,target)}_{\varepsilon}$ in the initial part of the total
circuit in Fig.~\ref{circuit}.
We define the following errors $\varepsilon$ on control and target qubits in every CNOT
gate: 
\begin{eqnarray}
{\bf CNOT}^{(A,B)}_{\varepsilon} \lvert \psi^{\bf{CNOT}}_{input} \rangle
&=&  \varepsilon_{A}\otimes\varepsilon_{B} \lvert \psi^{\bf{CNOT}}_{output} \rangle\\
 \varepsilon_{A}\otimes\varepsilon_{B} &=&\sum_{i=0}^{3} p_{i} 
\sigma_{A}^{i} \otimes \sum_{j=0}^{3}p_{j}\sigma_{B}^{j}. 
\end{eqnarray}
Here, $p_0 p_0 = 1-p = F$ 
and $p_i p_j = \frac{p}{15}$ except for
both $i=0$ and $j=0$.
$\sigma^{0},..,\sigma^{3}$ denote $\hat{I}$, $\hat{X}$, $\hat{Y}$, and
$\hat{Z}$ respectively.

We investigate the relation between the fidelity of the input
states and that of our output state.
Following the above setting, 
our initially shared seven Bell pairs $\lvert\psi_{init}^{\varepsilon,\bf{QNC}}\rangle$
include Pauli errors.
Each Bell pair, which is a combination of sixteen possible error
conditions, becomes a mixture of four states.
For example, we describe the state of Bell pair $AB$ below:
\begin{eqnarray}
{\bf CNOT}^{(A,B)}_{\varepsilon} H_{A} \lvert 00 \rangle_{AB} &=&  \varepsilon_{A}
\otimes \varepsilon_{B} \lvert \Psi^{+}_{AB} \rangle\\
&=& \left(1-\frac{4p}{5}\right) \lvert \Psi^{+}_{AB} \rangle 
+\frac{4p}{15} \lvert \Psi^{-}_{AB} \rangle \nonumber\\
&&+\frac{4p}{15} \lvert \Phi^{+}_{AB} \rangle 
+\frac{4p}{15} \lvert \Phi^{-}_{AB} \rangle.
\end{eqnarray}
This expression arises because of the symmetric effect of some errors
on Bell pairs, as in the following equations:
\begin{eqnarray}
\lvert \Psi^{+}_{AB} \rangle &=&
(\hat{I}_{A}\otimes\hat{I}_{B})\lvert \Psi^{+}_{AB} \rangle 
= (\hat{X}_{A}\otimes\hat{X}_{B})\lvert \Psi^{+}_{AB} \rangle \nonumber\\
&=&(\hat{Y}_{A}\otimes\hat{Y}_{B})\lvert \Psi^{+}_{AB} \rangle
=(\hat{Z}_{A}\otimes\hat{Z}_{B})\lvert \Psi^{+}_{AB} \rangle ,\\
\lvert \Phi^{+}_{AB} \rangle &=&
(\hat{I}_{A}\otimes\hat{X}_{B})\lvert \Psi^{+}_{AB} \rangle 
= (\hat{X}_{A}\otimes\hat{I}_{B})\lvert \Psi^{+}_{AB} \rangle \nonumber\\
&=&(\hat{Y}_{A}\otimes\hat{Z}_{B})\lvert \Psi^{+}_{AB} \rangle
=(\hat{Z}_{A}\otimes\hat{Y}_{B})\lvert \Psi^{+}_{AB} \rangle ,\\
\lvert \Psi^{-}_{AB} \rangle
 &=&(\hat{I}_{A}\otimes\hat{Z}_{B})\lvert \Psi^{+}_{AB} \rangle 
= (\hat{Z}_{A}\otimes\hat{I}_{B})\lvert \Psi^{+}_{AB} \rangle \nonumber\\
&=&(\hat{X}_{A}\otimes\hat{Y}_{B})\lvert \Psi^{+}_{AB} \rangle
=(\hat{Y}_{A}\otimes\hat{X}_{B})\lvert \Psi^{+}_{AB} \rangle ,\\
\lvert \Phi^{-}_{AB} \rangle &=&
(\hat{X}_{A}\otimes\hat{Z}_{B})\lvert \Psi^{+}_{AB} \rangle 
= (\hat{Z}_{A}\otimes\hat{X}_{B})\lvert \Psi^{+}_{AB} \rangle \nonumber\\
&=&(\hat{I}_{A}\otimes\hat{Y}_{B})\lvert \Psi^{+}_{AB} \rangle
=(\hat{Y}_{A}\otimes\hat{I}_{B})\lvert \Psi^{+}_{AB} \rangle .
\end{eqnarray}

Based on the above, we assume all Pauli errors exist on the target
qubits of CNOT gates in our initial resources. We show the relationship
between errors on initial states and final state in Table~\ref{epsilon_i}. 
For example, in the upper left corner of the table, the $\hat{I}_A \hat{X}_B$
entry indicates that an $X$ error on the initial  Bell pair $AB$
results in an error-free Bell pair $AF$ and an $X$ error on the 
Bell pair $BE$, so that the final state is $\ket{\Psi^+}_{AF}\ket{\Phi^+}_{BE}$.
\begin{table}[htbp]
\caption{\label{epsilon_i} The relationship between errors on initial Bell
  pairs and final states. Columns correspond to the type of errors on
  underbarred qubits of initial Bell pairs.}
\begin{ruledtabular}
 \begin{tabular}{llll}
Bell pair & $\hat{X}$& $\hat{Y}$& $\hat{Z}$\\
\hline
$\lvert \Psi^{+}\rangle_{A\underline{B}}$ & $\hat{I}_{A}\hat{X}_{B}$ &
$\hat{Z}_{A} \hat{X}_{B}$ & $\hat{Z}_{A} \hat{I}_{B}$ \\
$\lvert \Psi^{+}\rangle_{C\underline{D}}$ & $\hat{X}_{F}\hat{X}_{B}$ &
$\hat{Z}_{A} \hat{X}_{F} \hat{X}_{B}$ & $\hat{Z}_{A}  \hat{I}_{B}$ \\
$\lvert \Psi^{+}\rangle_{E\underline{F}}$ & $\hat{X}_{F}\hat{I}_{E}$ &
$\hat{X}_{F} \hat{Z}_{E}$ & $\hat{I}_{F} \hat{Z}_{E}$ \\
$\lvert \Psi^{+}\rangle_{G\underline{H}}$ & $\hat{X}_{F}\hat{X}_{B}$ &
$\hat{X}_{F} \hat{X}_{B}\hat{Z}_{E}$ & $\hat{I}_{F} \hat{Z}_{E}$ \\
$\lvert \Psi^{+}\rangle_{I\underline{J}}$ & $\hat{X}_{F}\hat{X}_{B}$ &
$\hat{Z}_{A} \hat{X}_{F} \hat{X}_{B} \hat{Z}_{E}$ & $\hat{Z}_{F} \hat{Z}_{E}$ \\
$\lvert \Psi^{+}\rangle_{K\underline{L}}$ & $\hat{I}_{A}\hat{X}_{B}$ &
$\hat{Z}_{A} \hat{X}_{B} \hat{Z}_{E}$ & $\hat{Z}_{A}  \hat{Z}_{E}$ \\
$\lvert \Psi^{+}\rangle_{M\underline{N}}$ & $\hat{X}_{F}\hat{I}_{B}$ &
$\hat{Z}_{A}\hat{X}_{F} \hat{Z}_{B}$ & $\hat{Z}_{A} \hat{Z}_{B}$ \\
 \end{tabular}
\end{ruledtabular}
\end{table}

\begin{figure}[htbp]
  \includegraphics[width=86mm]{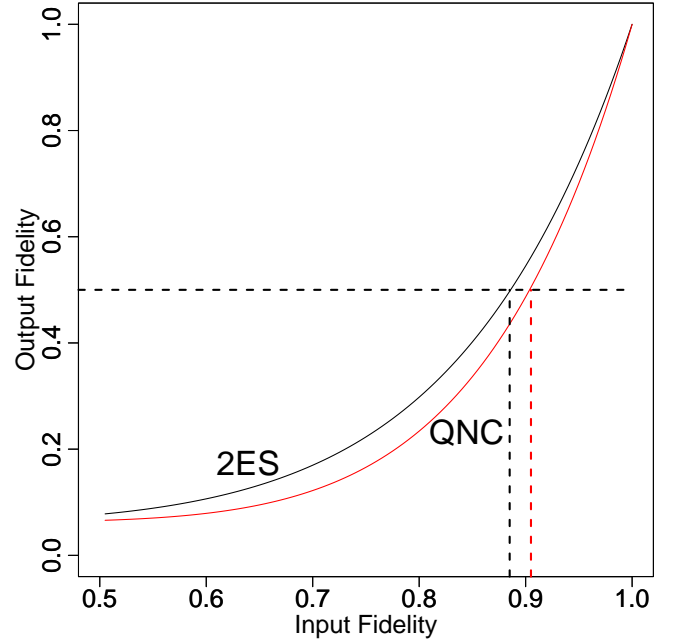}
 \caption{\label{SwappingvsQNC_Pauli}Joint  fidelity of the two output Bell
   pairs. We compare Swapping and QNC with general Pauli error model. Both
   show a substantial penalty compared to the fidelity of a single
   Bell pair.}
\end{figure}

We show the relationship between the
input fidelity and the output fidelity of our network coding protocol and
2-entanglement swapping in Fig.~\ref{SwappingvsQNC_Pauli}.
Here, the final state with $F_{output}<0.5$ has no practical use for
quantum communication.
When  $F_{input}\leq 0.88$, the 2ES protocol falls below $F_{out}=0.5$.
When  $F_{input}\leq 0.9$, the QNC protocol also falls below $F_{out}=0.5$.

\section{Incorporating gate errors}
\label{incorporating errors}
\begin{figure*}
\includegraphics[width=510pt]{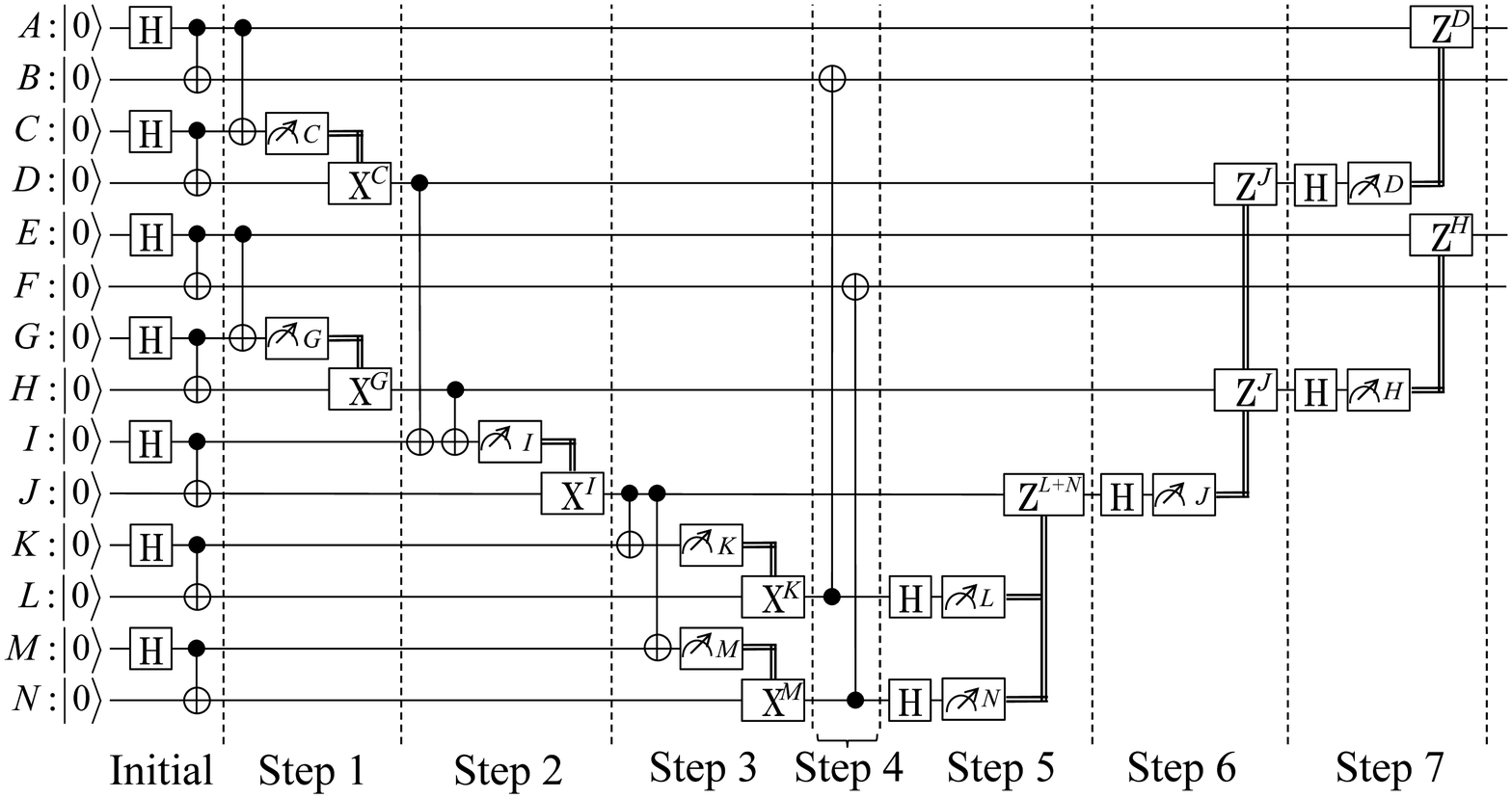}
\caption{\label{circuit}Complete circuit for QNC. Numbers refer
  to the step of QNC procedure. The initial Bell pair creation is
  modeled as a Hadamard gate followed by a CNOT, with a separate error
probability from the rest of the circuit.}
\end{figure*}

In this section,
we investigate the error propagation caused by local gates in
each encoding step which shown in Fig.~\ref{circuit}.
We introduce $\bf {Con_{\varepsilon}}$, $\bf {Add_{\varepsilon}}$,
$\bf {Fanout_{\varepsilon}}$, and $\bf {QNC_{\varepsilon}}$. 
These operators use $\bf{ CNOT_{\varepsilon}}$ within these operations.
Furthermore, the following error $\epsilon$ occur on
  all qubits in every measurement, single qubit gate, and waiting time.
\begin{eqnarray}
\epsilon =\sum^{3}_{i=0}p_i \sigma^{i}
\end{eqnarray}
Here, $p_0 = F$ and $p_i = \frac{p}{3}$ whenever $i \neq 0$.
In subsections~\ref{inc_1} through \ref{inc_5}, we give a step by
step qualitative analysis, then in subsection~\ref{inc_t} we present
the results of our Monte Carlo simulation of the complete circuit.
\subsection{Errors in Step 1}
\label{inc_1}
In step 1, the CNOT gate in Connection causes the following errors $\varepsilon^{(1)}$:
\begin{eqnarray}
{\bf Con}^{E}_{\varepsilon, G\rightarrow H}{\bf Con}^{A}_{\varepsilon,
  C\rightarrow D}\lvert \psi_{init} \rangle
= \varepsilon^{(1)} \lvert \psi_{final} \rangle .
\end{eqnarray}
When we assume the initial resources and CNOT gates in other steps do not
include errors, we can describe the relationship between errors in
this step and final states as shown in Table~\ref{epsilon_1}.

\begin{table}[htbp]
\caption{\label{epsilon_1} The relationship between errors caused by
  CNOT gates in Step 1 and final states. Columns correspond to the
  type of errors on underlined qubits.}
\begin{ruledtabular}
 \begin{tabular}{llll}
Qubit(underlined) & $\hat{X}$& $\hat{Y}$& $\hat{Z}$\\
\hline
$ \bf{CNOT}^{(\underline A ,C)}$ & $\hat{X}_{A}$ & $\hat{Y}_{A}$ &
$\hat{Z}_{A}$ \\
$ \bf{CNOT}^{(A,\underline{C})}$ & $\hat{X}_{B} \hat{X}_{F}$ &
$\hat{X}_{B} \hat{X}_{F}$ & $\hat{I}$ \\
$ \bf{CNOT}^{(\underline E,G)}$ & $\hat{X}_{E}$ & $\hat{Y}_{E}$ &
$\hat{Z}_{E}$ \\
$ \bf{CNOT}^{(E,\underline G)}$ & $\hat{X}_{B} \hat{X}_{F}$ &
$\hat{X}_{B} \hat{X}_{F}$ & $\hat{I}$ \\
 \end{tabular}
\end{ruledtabular}
\end{table}

\subsection{Errors in Step 2.}
In step 2, the CNOT gate in Add causes the following errors $\varepsilon^{(2)}$:
\begin{eqnarray}
{\bf Add}^{D,H}_{\varepsilon, I\rightarrow J} \lvert \psi_{(1)}\rangle
= \varepsilon^{(2)} \lvert \psi_{final} \rangle .
 \end{eqnarray}
When we assume the initial resources and CNOT gates in other steps do not
include errors, we can describe the relationship between errors in
this step and final states as shown in Table~\ref{epsilon_2}.

\begin{table}[htbp]
\caption{\label{epsilon_2} The relationship between errors caused by
  CNOT gates in Step 2 and final states. Columns correspond to the
  type of errors on underlined qubits.}
\begin{ruledtabular}
 \begin{tabular}{llll}
Qubit(underlined) & $\hat{X}$& $\hat{Y}$& $\hat{Z}$\\
\hline
$ \bf{CNOT}^{(\underline D ,I)}$ & $\hat{I}$ & $\hat{Z}_{A}$ &
$\hat{Z}_{A}$ \\
$ \bf{CNOT}^{(D,\underline I)}$ & $\hat{X}_{B} \hat{X}_{F}$ &
$\hat{X}_{B} \hat{Z}_{E} \hat{X}_{F}$ & $\hat{Z}_{E}$ \\
$ \bf{CNOT}^{(\underline H,I)}$ & $\hat{I}$ & $\hat{Z}_{E}$ &
$\hat{Z}_{E}$ \\
$ \bf{CNOT}^{(H,\underline I)}$ & $\hat{X}_{B} \hat{X}_{F}$ &
$\hat{X}_{B} \hat{X}_{F}$ & $\hat{I}$ \\
 \end{tabular}
\end{ruledtabular}
\end{table}

\subsection{Errors in Step 3.}
In step 3, the CNOT gate in Fanout causes the following errors $\varepsilon^{(3)}$:
\begin{eqnarray}
{\bf Fanout}^{J}_{\varepsilon, K\rightarrow L,M\rightarrow N} \lvert \psi_{(2)}\rangle 
= 
 \varepsilon^{(3)} \lvert \psi_{final} \rangle .
\end{eqnarray}
When we assume the initial resources and CNOT gates in other steps do not
include errors, we can describe the relationship between errors in
this step and final states as shown in Table~\ref{epsilon_3}.

\begin{table}[htbp]
\caption{\label{epsilon_3} The relationship between errors caused by
  CNOT gates in Step 3 and final states. Columns correspond to the
  type of errors on underlined qubits.}
\begin{ruledtabular}
 \begin{tabular}{llll}
Qubit(underlined) & $\hat{X}$& $\hat{Y}$& $\hat{Z}$\\
\hline
$ \bf{CNOT}^{(\underline J ,K)}$ & $\hat{X}_{F}$ & $\hat{Z}_{A}\hat{Z}_{E}\hat{X}_{F}$ &
$\hat{Z}_{A}\hat{Z}_{E}$ \\
$ \bf{CNOT}^{(J,\underline K)}$ & $\hat{X}_{B}$ &
$\hat{X}_{B}$ & $\hat{I}$ \\
$ \bf{CNOT}^{(\underline J,M)}$ & $\hat{I}$ & $\hat{Z}_{A}\hat{Z}_{E}$ &
$\hat{Z}_{A}\hat{Z}_{E}$ \\
$ \bf{CNOT}^{(J,\underline M)}$ & $\hat{X}_{F} $ &
$ \hat{X}_{F}$ & $\hat{I}$ \\
 \end{tabular}
\end{ruledtabular}
\end{table}

\subsection{Errors in Step 4.}
In step 4, the CNOT gate operations cause the following errors $\varepsilon^{(4)}$:
\begin{eqnarray}
{\bf CNOT}^{(N,F)}_{\varepsilon} {\bf CNOT}^{(L,B)}_{\varepsilon}
 \lvert \psi_{(3)}\rangle 
= 
\varepsilon^{(4)} \lvert \psi_{final} \rangle .
\end{eqnarray}
When we assume the initial resources and CNOT gates in other steps do not
include errors, we can describe the relationship between errors in
this step and final states as shown in Table~\ref{epsilon_4}.

\begin{table}
\caption{\label{epsilon_4} The relationship between errors caused by
  CNOT gates in Step 4 and final states. Columns correspond to the
  type of errors on underlined qubits.}
\begin{ruledtabular}
 \begin{tabular}{llll}
Qubit(underlined) & $\hat{X}$& $\hat{Y}$& $\hat{Z}$\\
\hline
$ \bf{CNOT}^{(\underline L ,B)}$ & $\hat{I}$ & $\hat{Z}_{A}\hat{Z}_{E}$& 
$ \hat{Z}_{A}\hat{Z}_{E}$ \\
$ \bf{CNOT}^{(L,\underline B)}$ & $\hat{X}_{B}$ &
$\hat{Y}_{B}$ & $\hat{Z}_{B}$ \\
$ \bf{CNOT}^{(\underline N,F)}$ & $\hat{I}$ & $\hat{Z}_{A}\hat{Z}_{E}$ &
$\hat{Z}_{A}\hat{Z}_{E}$ \\
$ \bf{CNOT}^{(N,\underline F)}$ & $\hat{X}_{F}$ &
$\hat{Y}_{F}$ & $\hat{Z}_{F}$ \\
 \end{tabular}
\end{ruledtabular}
\end{table}

\subsection{Errors in Step 5-7.}
\label{inc_5}
In these steps, no additional errors are added to the system.

\subsection{Simulations for total errors}
\label{inc_t}
\begin{figure}[htbp]
\centering
  \subfloat[]{\label{final_result_95}
  \includegraphics[clip,width=0.95\columnwidth]{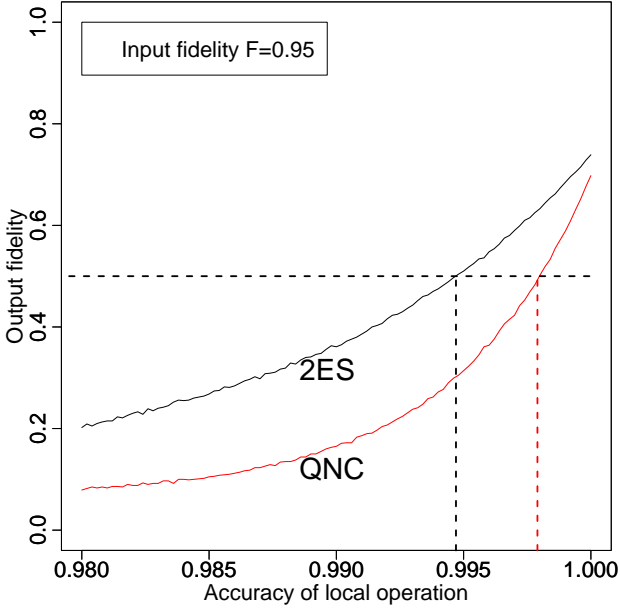}}
 \\
  \subfloat[]{\label{final_result_98}
  \includegraphics[clip,width=0.95\columnwidth]{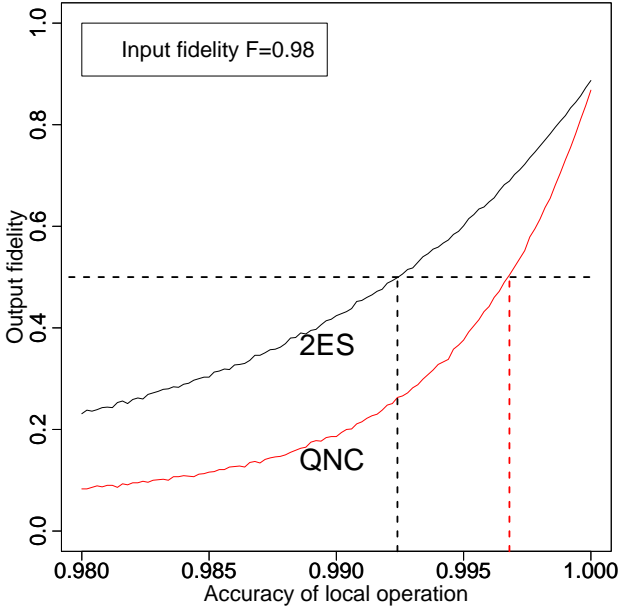}}
 \caption{\label{final_result}Comparison of Swapping and QNC with
   incorporating gate errors. Output fidelities correspond to the case
 with no error on either final Bell pair. (a) Initial fidelity
 $F=0.95$. (b) Initial fidelity $F=0.98$.} 
\end{figure}

Using these results, the final state can be described as follows:
\begin{eqnarray}
\bf{QNC}_{\varepsilon}\lvert\psi_{final}^{'''\bf{QNC}}\rangle
&=& \varepsilon^{(4)} \varepsilon^{(3)} \varepsilon^{(2)}
\varepsilon^{(1)} \varepsilon^{init} \lvert \psi_{final} \rangle \\
&=&\lvert \psi^{'''}_{final} \rangle
\end{eqnarray}
Then, we show the relation between the input fidelity of Bell pairs, the
accuracy of local operations, and the output fidelity in Fig.~\ref{final_result}.

To calculate these fidelities, we used Monte Carlo simulations.
In each simulation, the fidelitiies of Bell pairs are
fixed to $F=0.95$ or $F=0.98$. The accuracy of local operations is
changed from $F=0.980$ to $F=1.000$ using $\Delta F=0.001$. In
each parameter set, the simulation until we accumulate twenty thousand
errors on the final states (up to a maximum of one hundred million
times.).

\section{Conclusion}
\label{conclusion}
We have shown the propagation of errors in quantum network
coding protocols using the example of the butterfly network. We also
show the error threshold of quantum network coding in noisy quantum
repeater networks using Monte-Carlo simulations. 
We can see that QNC is more sensitive to local gate errors than
entanglement swapping. In the case of the butterfly network. 2ES
tolerates about twice the local error rate of QNC.
From these results, we see that each scheme is suitable for different purposes.
2ES is useful when the quantum resources are abundant or low
communication speed is permitted.
Quantum network coding is useful when the quantum resources are limited or high
communication speed is required.
The choice of scheme therefore depends on 
the environment of the quantum network and the quantum  application used.
We hope quantum network coding will be used in actual future repeater networks.

\section{Acknowledgements}
This work was supported by MEXT/JSPS KAKENHI Grant Number 25280034.

\bibliography{201504_PRA}

\begin{thebibliography}{29}%
\makeatletter
\providecommand \@ifxundefined [1]{%
 \@ifx{#1\undefined}
}%
\providecommand \@ifnum [1]{%
 \ifnum #1\expandafter \@firstoftwo
 \else \expandafter \@secondoftwo
 \fi
}%
\providecommand \@ifx [1]{%
 \ifx #1\expandafter \@firstoftwo
 \else \expandafter \@secondoftwo
 \fi
}%
\providecommand \natexlab [1]{#1}%
\providecommand \enquote  [1]{``#1''}%
\providecommand \bibnamefont  [1]{#1}%
\providecommand \bibfnamefont [1]{#1}%
\providecommand \citenamefont [1]{#1}%
\providecommand \href@noop [0]{\@secondoftwo}%
\providecommand \href [0]{\begingroup \@sanitize@url \@href}%
\providecommand \@href[1]{\@@startlink{#1}\@@href}%
\providecommand \@@href[1]{\endgroup#1\@@endlink}%
\providecommand \@sanitize@url [0]{\catcode `\\12\catcode `\$12\catcode
  `\&12\catcode `\#12\catcode `\^12\catcode `\_12\catcode `\%12\relax}%
\providecommand \@@startlink[1]{}%
\providecommand \@@endlink[0]{}%
\providecommand \url  [0]{\begingroup\@sanitize@url \@url }%
\providecommand \@url [1]{\endgroup\@href {#1}{\urlprefix }}%
\providecommand \urlprefix  [0]{URL }%
\providecommand \Eprint [0]{\href }%
\providecommand \doibase [0]{http://dx.doi.org/}%
\providecommand \selectlanguage [0]{\@gobble}%
\providecommand \bibinfo  [0]{\@secondoftwo}%
\providecommand \bibfield  [0]{\@secondoftwo}%
\providecommand \translation [1]{[#1]}%
\providecommand \BibitemOpen [0]{}%
\providecommand \bibitemStop [0]{}%
\providecommand \bibitemNoStop [0]{.\EOS\space}%
\providecommand \EOS [0]{\spacefactor3000\relax}%
\providecommand \BibitemShut  [1]{\csname bibitem#1\endcsname}%
\let\auto@bib@innerbib\@empty
\bibitem [{\citenamefont {Lloyd}\ \emph {et~al.}(2004)\citenamefont {Lloyd},
  \citenamefont {Shapiro}, \citenamefont {Wong}, \citenamefont {Kumar},
  \citenamefont {Shahriar},\ and\ \citenamefont {Yuen}}]{Lloyd_2004}%
  \BibitemOpen
  \bibfield  {author} {\bibinfo {author} {\bibfnamefont {S.}~\bibnamefont
  {Lloyd}}, \bibinfo {author} {\bibfnamefont {J.~H.}\ \bibnamefont {Shapiro}},
  \bibinfo {author} {\bibfnamefont {F.~N.~C.}\ \bibnamefont {Wong}}, \bibinfo
  {author} {\bibfnamefont {P.}~\bibnamefont {Kumar}}, \bibinfo {author}
  {\bibfnamefont {S.~M.}\ \bibnamefont {Shahriar}}, \ and\ \bibinfo {author}
  {\bibfnamefont {H.~P.}\ \bibnamefont {Yuen}},\ }\href {\doibase
  http://doi.acm.org/10.1145/1039111.1039118} {\bibfield  {journal} {\bibinfo
  {journal} {SIGCOMM Comput. Commun. Rev.}\ }\textbf {\bibinfo {volume} {34}},\
  \bibinfo {pages} {9} (\bibinfo {year} {2004})}\BibitemShut {NoStop}%
\bibitem [{\citenamefont {Kimble}(2008)}]{kimble2008quantum}%
  \BibitemOpen
  \bibfield  {author} {\bibinfo {author} {\bibfnamefont {H.~J.}\ \bibnamefont
  {Kimble}},\ }\href@noop {} {\bibfield  {journal} {\bibinfo  {journal}
  {Nature}\ }\textbf {\bibinfo {volume} {453}},\ \bibinfo {pages} {1023}
  (\bibinfo {year} {2008})}\BibitemShut {NoStop}%
\bibitem [{\citenamefont {Van{ }Meter}(2014)}]{van-meter14}%
  \BibitemOpen
  \bibfield  {author} {\bibinfo {author} {\bibfnamefont {R.}~\bibnamefont {Van{
  }Meter}},\ }\href@noop {} {\emph {\bibinfo {title} {Quantum Networking}}}\
  (\bibinfo  {publisher} {Wiley-ISTE},\ \bibinfo {year} {2014})\BibitemShut
  {NoStop}%
\bibitem [{\citenamefont {Bennett}\ and\ \citenamefont
  {Brassard}(1984)}]{bb84}%
  \BibitemOpen
  \bibfield  {author} {\bibinfo {author} {\bibfnamefont {C.~H.}\ \bibnamefont
  {Bennett}}\ and\ \bibinfo {author} {\bibfnamefont {G.}~\bibnamefont
  {Brassard}},\ }in\ \href@noop {} {\emph {\bibinfo {booktitle} {Proceedings of
  IEEE International Conference on Computers, Systems, and Signal
  Processing}}},\ Vol.~\bibinfo {volume} {11}\ (\bibinfo {year} {1984})\ pp.\
  \bibinfo {pages} {175--179}\BibitemShut {NoStop}%
\bibitem [{\citenamefont {Ekert}(1991)}]{qcb}%
  \BibitemOpen
  \bibfield  {author} {\bibinfo {author} {\bibfnamefont {A.}~\bibnamefont
  {Ekert}},\ }\href@noop {} {\bibfield  {journal} {\bibinfo  {journal} {Phys.
  Rev. Lett.}\ }\textbf {\bibinfo {volume} {67}},\ \bibinfo {pages} {661}
  (\bibinfo {year} {1991})}\BibitemShut {NoStop}%
\bibitem [{\citenamefont {Wootters}\ and\ \citenamefont
  {Zurek}(1982)}]{no_cloning}%
  \BibitemOpen
  \bibfield  {author} {\bibinfo {author} {\bibfnamefont {W.~K.}\ \bibnamefont
  {Wootters}}\ and\ \bibinfo {author} {\bibfnamefont {W.~H.}\ \bibnamefont
  {Zurek}},\ }\href@noop {} {\bibfield  {journal} {\bibinfo  {journal}
  {Nature}\ }\textbf {\bibinfo {volume} {299}},\ \bibinfo {pages} {802}
  (\bibinfo {year} {1982})}\BibitemShut {NoStop}%
\bibitem [{\citenamefont {Peev$}(2009)}]{SECOQC}%
  \BibitemOpen
  \bibfield  {author} {\bibinfo {author} {\bibfnamefont {M.}~\bibnamefont
  {Peev$}, \bibfnamefont {$~et~al}},\ }\href
  {http://stacks.iop.org/1367-2630/11/i=7/a=075001} {\bibfield  {journal}
  {\bibinfo  {journal} {New Journal of Physics}\ }\textbf {\bibinfo {volume}
  {11}},\ \bibinfo {pages} {075001} (\bibinfo {year} {2009})}\BibitemShut
  {NoStop}%
\bibitem [{\citenamefont {Sasaki$}(2011)}]{Sasaki_11}%
  \BibitemOpen
  \bibfield  {author} {\bibinfo {author} {\bibfnamefont {M.}~\bibnamefont
  {Sasaki$}, \bibfnamefont {$~et~al}},\ }\href {\doibase 10.1364/OE.19.010387}
  {\bibfield  {journal} {\bibinfo  {journal} {Opt. Express}\ }\textbf {\bibinfo
  {volume} {19}},\ \bibinfo {pages} {10387} (\bibinfo {year}
  {2011})}\BibitemShut {NoStop}%
\bibitem [{\citenamefont {Ben-Or}\ and\ \citenamefont
  {Hassidim}(2005)}]{byzantine}%
  \BibitemOpen
  \bibfield  {author} {\bibinfo {author} {\bibfnamefont {M.}~\bibnamefont
  {Ben-Or}}\ and\ \bibinfo {author} {\bibfnamefont {A.}~\bibnamefont
  {Hassidim}},\ }in\ \href@noop {} {\emph {\bibinfo {booktitle} {Proceedings of
  the thirty-seventh annual ACM symposium on Theory of computing}}}\ (\bibinfo
  {organization} {ACM},\ \bibinfo {year} {2005})\ pp.\ \bibinfo {pages}
  {481--485}\BibitemShut {NoStop}%
\bibitem [{\citenamefont {Broadbent}\ \emph {et~al.}(2010)\citenamefont
  {Broadbent}, \citenamefont {Fitzsimons},\ and\ \citenamefont
  {Kashefi}}]{broadbent2010measurement}%
  \BibitemOpen
  \bibfield  {author} {\bibinfo {author} {\bibfnamefont {A.}~\bibnamefont
  {Broadbent}}, \bibinfo {author} {\bibfnamefont {J.}~\bibnamefont
  {Fitzsimons}}, \ and\ \bibinfo {author} {\bibfnamefont {E.}~\bibnamefont
  {Kashefi}},\ }in\ \href {\doibase 10.1007/978-3-642-13678-8_2} {\emph
  {\bibinfo {booktitle} {Formal Methods for Quantitative Aspects of Programming
  Languages}}}\ (\bibinfo  {publisher} {Springer},\ \bibinfo {year} {2010})\
  pp.\ \bibinfo {pages} {43--86}\BibitemShut {NoStop}%
\bibitem [{\citenamefont {Briegel}\ \emph {et~al.}(1998)\citenamefont
  {Briegel}, \citenamefont {D\"ur}, \citenamefont {Cirac},\ and\ \citenamefont
  {Zoller}}]{repeater}%
  \BibitemOpen
  \bibfield  {author} {\bibinfo {author} {\bibfnamefont {H.-J.}\ \bibnamefont
  {Briegel}}, \bibinfo {author} {\bibfnamefont {W.}~\bibnamefont {D\"ur}},
  \bibinfo {author} {\bibfnamefont {J.~I.}\ \bibnamefont {Cirac}}, \ and\
  \bibinfo {author} {\bibfnamefont {P.}~\bibnamefont {Zoller}},\ }\href
  {\doibase 10.1103/PhysRevLett.81.5932} {\bibfield  {journal} {\bibinfo
  {journal} {Phys. Rev. Lett.}\ }\textbf {\bibinfo {volume} {81}},\ \bibinfo
  {pages} {5932} (\bibinfo {year} {1998})}\BibitemShut {NoStop}%
\bibitem [{\citenamefont {Duan}\ and\ \citenamefont
  {Monroe}(2010)}]{duan:RevModPhys.82.1209}%
  \BibitemOpen
  \bibfield  {author} {\bibinfo {author} {\bibfnamefont {L.-M.}\ \bibnamefont
  {Duan}}\ and\ \bibinfo {author} {\bibfnamefont {C.}~\bibnamefont {Monroe}},\
  }\href {\doibase 10.1103/RevModPhys.82.1209} {\bibfield  {journal} {\bibinfo
  {journal} {Rev. Mod. Phys.}\ }\textbf {\bibinfo {volume} {82}},\ \bibinfo
  {pages} {1209} (\bibinfo {year} {2010})}\BibitemShut {NoStop}%
\bibitem [{\citenamefont {Hucul}\ \emph {et~al.}(2014)\citenamefont {Hucul},
  \citenamefont {Inlek}, \citenamefont {Vittorini}, \citenamefont {Crocker},
  \citenamefont {Debnath}, \citenamefont {Clark},\ and\ \citenamefont
  {Monroe}}]{hucul2014modular}%
  \BibitemOpen
  \bibfield  {author} {\bibinfo {author} {\bibfnamefont {D.}~\bibnamefont
  {Hucul}}, \bibinfo {author} {\bibfnamefont {I.}~\bibnamefont {Inlek}},
  \bibinfo {author} {\bibfnamefont {G.}~\bibnamefont {Vittorini}}, \bibinfo
  {author} {\bibfnamefont {C.}~\bibnamefont {Crocker}}, \bibinfo {author}
  {\bibfnamefont {S.}~\bibnamefont {Debnath}}, \bibinfo {author} {\bibfnamefont
  {S.}~\bibnamefont {Clark}}, \ and\ \bibinfo {author} {\bibfnamefont
  {C.}~\bibnamefont {Monroe}},\ }\href@noop {} {\bibfield  {journal} {\bibinfo
  {journal} {Nature Physics}\ }\textbf {\bibinfo {volume} {11}},\ \bibinfo
  {pages} {37} (\bibinfo {year} {2014})}\BibitemShut {NoStop}%
\bibitem [{\citenamefont {D\"ur}\ and\ \citenamefont
  {Briegel}(2007)}]{Briegel_2007}%
  \BibitemOpen
  \bibfield  {author} {\bibinfo {author} {\bibfnamefont {W.}~\bibnamefont
  {D\"ur}}\ and\ \bibinfo {author} {\bibfnamefont {H.~J.}\ \bibnamefont
  {Briegel}},\ }\href@noop {} {\bibfield  {journal} {\bibinfo  {journal}
  {Reports on Progress in Physics}\ }\textbf {\bibinfo {volume} {70}},\
  \bibinfo {pages} {1381} (\bibinfo {year} {2007})}\BibitemShut {NoStop}%
\bibitem [{\citenamefont {Van~Meter}\ \emph {et~al.}(2009)\citenamefont
  {Van~Meter}, \citenamefont {Ladd}, \citenamefont {Munro},\ and\ \citenamefont
  {Nemoto}}]{VanMeter_2009_2}%
  \BibitemOpen
  \bibfield  {author} {\bibinfo {author} {\bibfnamefont {R.}~\bibnamefont
  {Van~Meter}}, \bibinfo {author} {\bibfnamefont {T.~D.}\ \bibnamefont {Ladd}},
  \bibinfo {author} {\bibfnamefont {W.~J.}\ \bibnamefont {Munro}}, \ and\
  \bibinfo {author} {\bibfnamefont {K.}~\bibnamefont {Nemoto}},\ }\href
  {http://dblp.uni-trier.de/db/journals/ton/ton17.html#MeterLMN09} {\bibfield
  {journal} {\bibinfo  {journal} {IEEE/ACM Trans. Netw.}\ }\textbf {\bibinfo
  {volume} {17}},\ \bibinfo {pages} {1002} (\bibinfo {year}
  {2009})}\BibitemShut {NoStop}%
\bibitem [{\citenamefont {Jiang}\ \emph {et~al.}(2009)\citenamefont {Jiang},
  \citenamefont {Taylor}, \citenamefont {Nemoto}, \citenamefont {Munro},
  \citenamefont {Van~Meter},\ and\ \citenamefont {Lukin}}]{VanMeter_2009}%
  \BibitemOpen
  \bibfield  {author} {\bibinfo {author} {\bibfnamefont {L.}~\bibnamefont
  {Jiang}}, \bibinfo {author} {\bibfnamefont {J.~M.}\ \bibnamefont {Taylor}},
  \bibinfo {author} {\bibfnamefont {K.}~\bibnamefont {Nemoto}}, \bibinfo
  {author} {\bibfnamefont {W.~J.}\ \bibnamefont {Munro}}, \bibinfo {author}
  {\bibfnamefont {R.}~\bibnamefont {Van~Meter}}, \ and\ \bibinfo {author}
  {\bibfnamefont {M.~D.}\ \bibnamefont {Lukin}},\ }\href {\doibase
  10.1103/PhysRevA.79.032325} {\bibfield  {journal} {\bibinfo  {journal} {Phys.
  Rev. A}\ }\textbf {\bibinfo {volume} {79}},\ \bibinfo {pages} {032325}
  (\bibinfo {year} {2009})}\BibitemShut {NoStop}%
\bibitem [{\citenamefont {Munro}\ \emph {et~al.}(2010)\citenamefont {Munro},
  \citenamefont {Harrison}, \citenamefont {Stephens}, \citenamefont {Devitt},\
  and\ \citenamefont {Nemoto}}]{Munro_2010}%
  \BibitemOpen
  \bibfield  {author} {\bibinfo {author} {\bibfnamefont {W.~J.}\ \bibnamefont
  {Munro}}, \bibinfo {author} {\bibfnamefont {K.~A.}\ \bibnamefont {Harrison}},
  \bibinfo {author} {\bibfnamefont {A.~M.}\ \bibnamefont {Stephens}}, \bibinfo
  {author} {\bibfnamefont {S.~J.}\ \bibnamefont {Devitt}}, \ and\ \bibinfo
  {author} {\bibfnamefont {K.}~\bibnamefont {Nemoto}},\ }\href {\doibase
  10.1038/nphoton.2010.213} {\bibfield  {journal} {\bibinfo  {journal} {Nature
  Photonics}\ }\textbf {\bibinfo {volume} {4}},\ \bibinfo {pages} {792}
  (\bibinfo {year} {2010})}\BibitemShut {NoStop}%
\bibitem [{\citenamefont {Fowler}\ \emph {et~al.}(2010)\citenamefont {Fowler},
  \citenamefont {Wang}, \citenamefont {Hill}, \citenamefont {Ladd},
  \citenamefont {Van~Meter},\ and\ \citenamefont {Hollenberg}}]{VanMeter_2010}%
  \BibitemOpen
  \bibfield  {author} {\bibinfo {author} {\bibfnamefont {A.~G.}\ \bibnamefont
  {Fowler}}, \bibinfo {author} {\bibfnamefont {D.~S.}\ \bibnamefont {Wang}},
  \bibinfo {author} {\bibfnamefont {C.~D.}\ \bibnamefont {Hill}}, \bibinfo
  {author} {\bibfnamefont {T.~D.}\ \bibnamefont {Ladd}}, \bibinfo {author}
  {\bibfnamefont {R.}~\bibnamefont {Van~Meter}}, \ and\ \bibinfo {author}
  {\bibfnamefont {L.~C.~L.}\ \bibnamefont {Hollenberg}},\ }\href@noop {}
  {\bibfield  {journal} {\bibinfo  {journal} {Phys. Rev. Lett.}\ }\textbf
  {\bibinfo {volume} {104}},\ \bibinfo {pages} {180503} (\bibinfo {year}
  {2010})}\BibitemShut {NoStop}%
\bibitem [{\citenamefont {\ifmmode~\dot{Z}\else \.{Z}\fi{}ukowski}\ \emph
  {et~al.}(1993)\citenamefont {\ifmmode~\dot{Z}\else \.{Z}\fi{}ukowski},
  \citenamefont {Zeilinger}, \citenamefont {Horne},\ and\ \citenamefont
  {Ekert}}]{entanglement_swapping}%
  \BibitemOpen
  \bibfield  {author} {\bibinfo {author} {\bibfnamefont {M.}~\bibnamefont
  {\ifmmode~\dot{Z}\else \.{Z}\fi{}ukowski}}, \bibinfo {author} {\bibfnamefont
  {A.}~\bibnamefont {Zeilinger}}, \bibinfo {author} {\bibfnamefont {M.~A.}\
  \bibnamefont {Horne}}, \ and\ \bibinfo {author} {\bibfnamefont {A.~K.}\
  \bibnamefont {Ekert}},\ }\href {\doibase 10.1103/PhysRevLett.71.4287}
  {\bibfield  {journal} {\bibinfo  {journal} {Phys. Rev. Lett.}\ }\textbf
  {\bibinfo {volume} {71}},\ \bibinfo {pages} {4287} (\bibinfo {year}
  {1993})}\BibitemShut {NoStop}%
\bibitem [{\citenamefont {Ahlswede}\ \emph {et~al.}(2000)\citenamefont
  {Ahlswede}, \citenamefont {Cai}, \citenamefont {Li},\ and\ \citenamefont
  {Yeung}}]{network_coding}%
  \BibitemOpen
  \bibfield  {author} {\bibinfo {author} {\bibfnamefont {R.}~\bibnamefont
  {Ahlswede}}, \bibinfo {author} {\bibfnamefont {N.}~\bibnamefont {Cai}},
  \bibinfo {author} {\bibfnamefont {S.~R.}\ \bibnamefont {Li}}, \ and\ \bibinfo
  {author} {\bibfnamefont {R.~W.}\ \bibnamefont {Yeung}},\ }\href@noop {}
  {\bibfield  {journal} {\bibinfo  {journal} {IEEE Transactions on Information
  Theory}\ }\textbf {\bibinfo {volume} {46}},\ \bibinfo {pages} {1204}
  (\bibinfo {year} {2000})}\BibitemShut {NoStop}%
\bibitem [{\citenamefont {Hayashi}\ \emph {et~al.}(2007)\citenamefont
  {Hayashi}, \citenamefont {Iwama}, \citenamefont {Nishimura}, \citenamefont
  {Raymond},\ and\ \citenamefont {Yamashita}}]{quantum_coding}%
  \BibitemOpen
  \bibfield  {author} {\bibinfo {author} {\bibfnamefont {M.}~\bibnamefont
  {Hayashi}}, \bibinfo {author} {\bibfnamefont {K.}~\bibnamefont {Iwama}},
  \bibinfo {author} {\bibfnamefont {H.}~\bibnamefont {Nishimura}}, \bibinfo
  {author} {\bibfnamefont {R.}~\bibnamefont {Raymond}}, \ and\ \bibinfo
  {author} {\bibfnamefont {S.}~\bibnamefont {Yamashita}},\ }in\ \href@noop {}
  {\emph {\bibinfo {booktitle} {Symposium on Theoretical Aspects of Computer
  Science}}}\ (\bibinfo {year} {2007})\ pp.\ \bibinfo {pages}
  {610--621}\BibitemShut {NoStop}%
\bibitem [{\citenamefont {Leung}\ \emph {et~al.}(2010)\citenamefont {Leung},
  \citenamefont {Oppenheim},\ and\ \citenamefont {Winter}}]{quantum_coding2}%
  \BibitemOpen
  \bibfield  {author} {\bibinfo {author} {\bibfnamefont {D.}~\bibnamefont
  {Leung}}, \bibinfo {author} {\bibfnamefont {J.}~\bibnamefont {Oppenheim}}, \
  and\ \bibinfo {author} {\bibfnamefont {A.}~\bibnamefont {Winter}},\
  }\href@noop {} {\bibfield  {journal} {\bibinfo  {journal} {IEEE Transactions
  on Information Theory}\ }\textbf {\bibinfo {volume} {56}},\ \bibinfo {pages}
  {3478} (\bibinfo {year} {2010})}\BibitemShut {NoStop}%
\bibitem [{\citenamefont {Hayashi}(2007)}]{quantum_coding3}%
  \BibitemOpen
  \bibfield  {author} {\bibinfo {author} {\bibfnamefont {M.}~\bibnamefont
  {Hayashi}},\ }\href {\doibase 10.1103/PhysRevA.76.040301} {\bibfield
  {journal} {\bibinfo  {journal} {Phys. Rev. A}\ }\textbf {\bibinfo {volume}
  {76}},\ \bibinfo {pages} {040301} (\bibinfo {year} {2007})}\BibitemShut
  {NoStop}%
\bibitem [{\citenamefont {Shi}\ and\ \citenamefont
  {Soljanin}(2006)}]{quantum_coding4}%
  \BibitemOpen
  \bibfield  {author} {\bibinfo {author} {\bibfnamefont {Y.}~\bibnamefont
  {Shi}}\ and\ \bibinfo {author} {\bibfnamefont {E.}~\bibnamefont {Soljanin}},\
  }in\ \href@noop {} {\emph {\bibinfo {booktitle} {40th Annual Conference on
  Information Sciences and Systems}}}\ (\bibinfo {year} {2006})\ pp.\ \bibinfo
  {pages} {871--876}\BibitemShut {NoStop}%
\bibitem [{\citenamefont {Kobayashi}\ \emph {et~al.}(2009)\citenamefont
  {Kobayashi}, \citenamefont {Le~Gall}, \citenamefont {Nishimura},\ and\
  \citenamefont {Roetteler}}]{kobayashi_coding}%
  \BibitemOpen
  \bibfield  {author} {\bibinfo {author} {\bibfnamefont {H.}~\bibnamefont
  {Kobayashi}}, \bibinfo {author} {\bibfnamefont {F.}~\bibnamefont {Le~Gall}},
  \bibinfo {author} {\bibfnamefont {H.}~\bibnamefont {Nishimura}}, \ and\
  \bibinfo {author} {\bibfnamefont {M.}~\bibnamefont {Roetteler}},\ }in\
  \href@noop {} {\emph {\bibinfo {booktitle} {36th International Colloquium on
  Automata, Languages and Programming}}}\ (\bibinfo {year} {2009})\ pp.\
  \bibinfo {pages} {622--633}\BibitemShut {NoStop}%
\bibitem [{\citenamefont {Kobayashi}\ \emph {et~al.}(2010)\citenamefont
  {Kobayashi}, \citenamefont {Le~Gall}, \citenamefont {Nishimura},\ and\
  \citenamefont {Roetteler}}]{kobayashi_coding2}%
  \BibitemOpen
  \bibfield  {author} {\bibinfo {author} {\bibfnamefont {H.}~\bibnamefont
  {Kobayashi}}, \bibinfo {author} {\bibfnamefont {F.}~\bibnamefont {Le~Gall}},
  \bibinfo {author} {\bibfnamefont {H.}~\bibnamefont {Nishimura}}, \ and\
  \bibinfo {author} {\bibfnamefont {M.}~\bibnamefont {Roetteler}},\ }in\
  \href@noop {} {\emph {\bibinfo {booktitle} {2010 IEEE International Symposium
  on Information Theory}}}\ (\bibinfo {year} {2010})\ pp.\ \bibinfo {pages}
  {2686--2690}\BibitemShut {NoStop}%
\bibitem [{\citenamefont {Kobayashi}\ \emph {et~al.}(2011)\citenamefont
  {Kobayashi}, \citenamefont {Le~Gall}, \citenamefont {Nishimura},\ and\
  \citenamefont {Roetteler}}]{kobayashi_coding3}%
  \BibitemOpen
  \bibfield  {author} {\bibinfo {author} {\bibfnamefont {H.}~\bibnamefont
  {Kobayashi}}, \bibinfo {author} {\bibfnamefont {F.}~\bibnamefont {Le~Gall}},
  \bibinfo {author} {\bibfnamefont {H.}~\bibnamefont {Nishimura}}, \ and\
  \bibinfo {author} {\bibfnamefont {M.}~\bibnamefont {Roetteler}},\ }in\
  \href@noop {} {\emph {\bibinfo {booktitle} {2011 IEEE International Symposium
  on Information Theory}}}\ (\bibinfo {year} {2011})\ pp.\ \bibinfo {pages}
  {109--113}\BibitemShut {NoStop}%
\bibitem [{\citenamefont {Satoh}\ \emph {et~al.}(2012)\citenamefont {Satoh},
  \citenamefont {Le~Gall},\ and\ \citenamefont {Imai}}]{repeater_coding}%
  \BibitemOpen
  \bibfield  {author} {\bibinfo {author} {\bibfnamefont {T.}~\bibnamefont
  {Satoh}}, \bibinfo {author} {\bibfnamefont {F.}~\bibnamefont {Le~Gall}}, \
  and\ \bibinfo {author} {\bibfnamefont {H.}~\bibnamefont {Imai}},\ }\href
  {\doibase 10.1103/PhysRevA.86.032331} {\bibfield  {journal} {\bibinfo
  {journal} {Phys. Rev. A}\ }\textbf {\bibinfo {volume} {86}},\ \bibinfo
  {pages} {032331} (\bibinfo {year} {2012})}\BibitemShut {NoStop}%
\bibitem [{\citenamefont {Aparicio}\ and\ \citenamefont {Van{
  }Meter}(2011)}]{repeater-muxing}%
  \BibitemOpen
  \bibfield  {author} {\bibinfo {author} {\bibfnamefont {L.}~\bibnamefont
  {Aparicio}}\ and\ \bibinfo {author} {\bibfnamefont {R.}~\bibnamefont {Van{
  }Meter}},\ }in\ \href@noop {} {\emph {\bibinfo {booktitle} {Proc. SPIE}}},\
  Vol.\ \bibinfo {volume} {8163}\ (\bibinfo {year} {2011})\ p.\ \bibinfo
  {pages} {816308}\BibitemShut {NoStop}%
\end{thebibliography}%

\end{document}